\documentclass[useAMS]{mn2e}
\usepackage{graphicx}
\begin{document}
\def\simlt{\mathrel{\rlap{\lower 3pt\hbox{$\sim$}}
        \raise 2.0pt\hbox{$<$}}}
\def\simgt{\mathrel{\rlap{\lower 3pt\hbox{$\sim$}}
        \raise 2.0pt\hbox{$>$}}}
\def\bj{b_{\rm\scriptscriptstyle J}}
\def\rt{r_{\rm\scriptscriptstyle T}}
\def\rp{r_{\rm\scriptscriptstyle P}}
\renewcommand{\labelenumi}{(\arabic{enumi})}
\title[The radio properties of optically obscured {\it Spitzer} sources ]
{The radio properties of optically obscured {\it Spitzer} sources}
\author[Manuela Magliocchetti et al.]
{\parbox[t]\textwidth{M. Magliocchetti$^{1,2,3}$, P. Andreani$^{2,1}$, M. A. Zwaan$^{2}$}\\
\tt $^1$ INAF, Osservatorio Astronomico di Trieste, Via Tiepolo 11, 34100,
Trieste, Italy\\
\tt $^2$ ESO, Karl-Schwarzschild-Str.2, D-85748, Garching, Germany\\
\tt $^3$ SISSA, Via Beirut 4, 34014, Trieste, Italy\\}
\maketitle
\begin{abstract}
This paper analyses the radio properties of a subsample of 
optically obscured ($R\ge 25.5$) galaxies observed at 24$\mu$m by the 
{\it Spitzer Space Telescope} within the First Look Survey. 
96 $F_{24\mu\rm m}\ge 0.35$~mJy  objects out of 510
are found to have a radio counterpart at 1.4~GHz, 610~MHz 
or at both frequencies respectively down to $\sim 40\mu$Jy and 
$\sim 200\mu$Jy. 
IRAC photometry sets the majority of them in the redshift interval $z\simeq [1-3]$ 
and allows for a broad distinction between 
AGN-dominated galaxies ($\sim 47$\% of the radio-identified sample) 
and systems powered by intense star-formation ($\sim 13$\%), the remaining 
objects being impossible to classify. The percentage of radio 
identifications is a strong function of 24$\mu$m flux: almost 
all sources brighter than $F_{24\mu\rm m}\sim 2$~mJy are endowed with a 
radio flux at both 1.4~GHz and 610~MHz, while this fraction drastically 
decreases by lowering the 24$\mu$m flux level. The radio number counts at both 
radio frequencies suggest that the physical process(es) responsible for radio 
activity in these objects have a common origin regardless of whether the 
source shows mid-IR emission compatible with being an obscured AGN or a 
star-forming galaxy. We also find that both candidate AGN and star-forming 
systems follow (although with a large scatter) the relationship between 
1.4~GHz and 24$\mu$m fluxes reported by Appleton et al. (2004) which 
identifies sources undergoing intense star formation activity. However, a more 
scattered relation is observed between 24$\mu$m and 610~MHz fluxes. On the other hand, 
the inferred radio spectral indices $\alpha$ indicate that a large fraction
of objects in our sample ($\sim 60$\% of all galaxies with estimated $\alpha$) 
may belong to the population of Ultra Steep Spectrum (USS) Sources, 
typically 'frustrated' radio-loud AGN.  
We interpret our findings as a strong indication for concurrent 
AGN and star-forming activity,
whereby the 1.4~GHz flux is of thermal origin, while 
that at 610~GHz mainly stems from the nuclear source.
\end{abstract}

\section{Introduction}
The advent of the {\it Spitzer Space Telescope} has marked a fundamental 
milestone in our understanding of the assembly history of massive spheroidal 
galaxies, one of the major issues for galaxy formation models. 
The unprecedented sensitivity of the Multiband Imaging Photometer for {\it Spitzer} (MIPS) 
at 24$\mu$m has in fact for the first time allowed the detection at high redshifts of a 
population of Luminous and UltraLuminous Infrared Galaxies (LIRGs; ULIRGs) with huge 
infrared luminosities ($L_{\rm IR}>10^{11}L_{\odot}$). 
Such sources are underluminous at rest frame optical and UV wavelengths because they are 
reprocessing and radiating much of their energy in the IR (e.g. Sanders \& Mirabel 1996). 
As a consequence, LIRGs and ULIRGs at high redshifts had been missed so far either due to 
their extreme optical faintness or because previous infrared missions such as the 
InfraRed Astronomical Satellite (IRAS) or the Infrared Space Observatory (ISO) did 
not have enough sensitivity to push the observations beyond $z \sim 1$.

Recent studies have shown that these objects, while relatively rare in the local universe 
(e.g. Sanders \& Mirabel 1996), become an increasingly significant population at higher 
redshifts (e.g. Le Floc'h et al. 2004; 2005; Lonsdale et al. 2004; Caputi et al. 2006; 
2007) and likely dominate the luminosity density at $z>1$ 
(see e.g. Dole et al. 2006). Their space density, found to range between $10^{-4}$ and a 
few $10^{-5}$~Mpc$^{-3}$ according to the selection criteria adopted by different studies 
(see e.g. Caputi et al 2007; Daddi et al. 2007a, Magliocchetti et al. 2007a), is 
a factor of 10 to 100 higher than that of optically selected quasars in the same 
redshift range (e.g. Porciani, Magliocchetti \& Norberg 2004). 
Furthermore, clustering studies (e.g. Magliocchetti 
et al. 2007; 2007a; Farrah et al. 2006) prove that, at variance with their local 
counterparts, LIRGs and ULIRGs at $z\sim 2$ are associated with extremely massive 
($M\simgt 10^{13} M_{\odot}$, where $M$ here refers to the dark matter) structures, 
only second to those which locally host very rich clusters of galaxies.
Given their properties, it then appears clear that these sources represent a fundamental 
phase in the build up of massive galactic bulges, and in the growth of their supermassive 
black holes.

Emission line diagnostics for bright ($F_{24\mu \rm m}\simgt 1$~mJy) 
mid-IR samples of LIRGs and ULIRGs at $z\sim 2$ in the near and mid-IR spectral regimes 
(see e.g. Yan et al. 2005; 2007; Weedman et al. 2006; 2006a; Brand et al. 2007; 
Martinez-Sansigre et al. 2006a)
have shown these sources to be a mixture of obscured type1-type2 AGN and systems 
undergoing intense star formation activity. These findings are confirmed by 
photometric follow up mainly undertaken in the mid-IR and X-ray (both soft and hard) 
bands which also prove that the fraction of galaxies dominated by a contribution of AGN 
origin is drastically reduced at faint mid-IR fluxes (e.g. Brand et al. 2006; 
Weedman et al. 2006;  Treister et al. 2006; Magliocchetti et al. 2007).
Unfortunately, the exact proportion of AGN vs starforming dominated galaxies is still 
undetermined. This separation is further complicated by the existence of a 
noticeable number of mixed systems where both star formation and AGN activity 
significantly contribute to the IR emission. For instance, Daddi et al. (2007a) find that 
about 20\% of 24$\mu$m-selected galaxies in the GOODS sample show a mid-IR excess which 
is not possible to reconcile with pure star-forming activity. Such a fraction increases 
to $\sim 50-60$\% at the highest (stellar) masses probed by their study.

Clearly, understanding how the mid-IR sources divide between starbursts, 
AGN and composite systems is now the next essential step in order 
understand the relationships amongst the formation and evolution of stars, galaxies 
and massive black holes powering AGNs within dusty environments and more generally 
within massive systems observed at the peak of their activity.

This paper approaches the study of the population of optically faint luminous 
infrared galaxies from the point of view of their multifrequency radio emission. 
Diagnostics based on the radio signal steming from these sources can in fact 
provide precious information on the process(es) which are actively taking place 
within such systems. Enhanced radio activity can stem from supernova remnants 
associated with regions which are vigorously forming stars, 
or originate from nuclear activity (AGN-dominated sources). These two processes determine a rather 
different spectral behaviour at radio wavelengths: star-forming systems are in general 
characterized by radio spectra which feature power-law shapes with slope 
(hereafter called {\it radio spectral index} $\alpha$, defined as $F\propto \nu^{-\alpha}$, 
with $F$ radio flux and $\nu$ radio frequency) of the order 
of $\sim$0.7-0.8, while typical radio-loud quasars exhibit values for $\alpha$ between 0 and 
0.5 even though, especially at high redshifts, there is a non negligible population 
of radio-loud sources with very high, $\alpha>1$, values (see e.g. De Breuck et al. 
2000). 

Radio counterparts to the 
510 optically faint mid-IR selected sources drawn from the whole First Look Survey 
sample (Fadda et al. 2006) have been searched in the overlapping region between MIPS and 
IRAC observations 
(Magliocchetti et al. 2007). Despite not having direct redshift estimates 
except for a handful of cases (e.g. Weedman et al. 2006; Yan et al. 
2005; 2007; Martinez-Sansigre et al. 2006a), mid-IR photometry indicates that 
the overwhelming majority of such sources reside at redshifts $1.7\simlt z\simlt 2.5$ 
(\S~4; see also Brand et al. 2007; Houck et al. 2005; Weedman et al. 2006a).\\
The First Look Survey region provides an excellent laboratory for investigations of 
the radio properties of dusty galaxies set at redshifts $z\sim 2$ as its area has 
been observed at a number of radio frequencies down to very low flux densities 
(Condon et al. 2003; Morganti et al. 2004; Garn et al. 2007), therefore maximizing 
our chances of finding radio emitting galaxies.  

The layout of the paper is as follows. In \S2 we introduce the
parent (mid-IR and radio) catalogues, while in \S3 we present 
the matching procedure leading to  
the sample of optically obscured {\it Spitzer}-selected sources with 
a radio counterpart at 1.4~GHz and/or 610~MHz.
\S4 uses IRAC photometry to provide some information on the typology of 
such objects (i.e. whether mainly powered by an obscured AGN or by a 
starburst) and also -- where possible -- to assign them to a redshift interval. 
\S5 presents the results on the radio number counts both at 1.4~GHz and 
610~MHz (\S5.1) and on the relationship between radio and 24$\mu$m 
emission (\S5.2). \S6 discusses our findings on the 1.4~GHz vs 610~MHz 
radio spectral indices for the objects in our sample, while \S7 summarizes 
our conclusions.

\section{Parent Catalogues}
\subsection{Optically Obscured {\it Spitzer} sources}
\begin{figure}
\includegraphics[width=8.0cm]{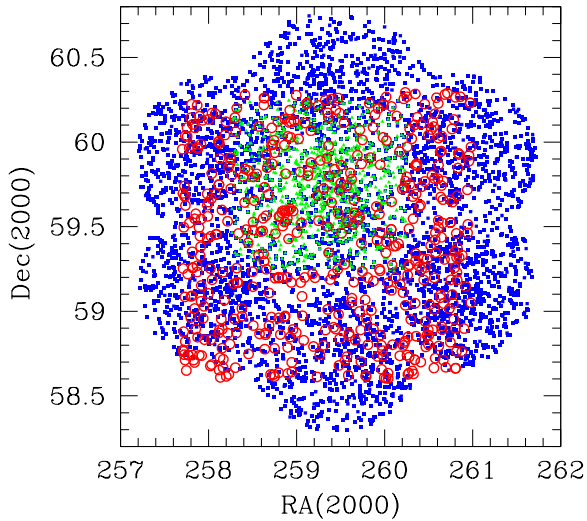} 
\caption{Sky distribution of sources in the FLS area. The (blue) squares are 
radio objects from the 610~MHz GMRT observations (Garn et al. 2007), the (green) triangles 
are the 1.4~GHz Morganti et al. (2004) sources, while the (red) open circles are 
the 510, 24$\mu$m-selected sources brighter than 0.35~mJy without an 
optical counterpart brighter than $R\simeq 25.5$ coming from the Magliocchetti et al. 
(2007) sample (see text for detail).
\label{fig:sky}}
\end{figure}

The primary selection of the sources in this work 
comes from {\it Spitzer}-MIPS 24$\mu$m observations of the $\sim 4$ square 
degrees region denoted as First-Look Survey  
(FLS, Fadda et al. 2006). Out of the original parent catalogue, 
in the 2.85 square degrees area covered by both MIPS and IRAC 
(Lacy et al. 2005) data, Magliocchetti et al. (2007) selected sources 
with 24$\mu$m fluxes brighter than 
0.35~mJy, limit which ensures $\sim 100$\% completeness of the MIPS dataset. 
As a further, crucial selection, Magliocchetti et al. (2007) required 
the above objects to be optically obscured. In practical terms, 
the requirement was that these sources had to be fainter than $R\sim 25.5$ 
(i.e. undetected) in the 
KPNO catalogue (Fadda et al. 2004) which covers the entire FLS region. 
As indicated by colour-colour evolutionary tracks for a number of known templates 
(see e.g. Figure 2 of the Magliocchetti et al. 2007 paper), and recently confirmed 
by results based on both mid-IR and near-IR spectroscopy, 
the above constraint forces the overwhelming majority of these objects 
to reside at redshifts beyond $z\sim 1.6$, with an average of $z\sim 2$ 
(see e.g. Weedman et al. 2006; Yan et al. 2005; 2007; Brand et al. 2007). 
The final number of obscured, $F_{24\mu\rm m}\ge 0.35$~mJy, 
galaxies in the overlapping MIPS-IRAC region is 510. 
Diagnostics based on the 8$\mu$m/24$\mu$m flux density ratios indicate that these 
are a mixture of galaxies undergoing an extreme event of stellar formation and 
obscured AGN, this latter population dominating at bright, 
$F_{24 \mu m}\simgt 0.8$~mJy fluxes (see also Brand et al. 2006; Treister et al. 2006).\\ 
The distribution of all the 510 optically obscured, 24$\mu$m-selected sources 
as obtained by Magliocchetti et al. (2007) is represented by the open (red) 
circles in Figure~\ref{fig:sky}.

\subsection{The Radio Catalogues}
The whole FLS region was observed at 1.4~GHz by 
Condon et al. (2003). They used the B configuration of the VLA to obtain radio 
images with $\sigma_F\sim 23 \mu\rm Jy$ beam$^{-1}$ rms fluctuations, 
$\theta=5^{\prime\prime}$ resolution and a positional rms of about 
$0.5$ arcsec. The resulting catalogue contains 3565 radio sources with 
peak flux densities $S_p\ge 5\sigma_F=115 \mu$Jy beam$^{-1}$.

A smaller, $\sim$1~deg$^2$ area of the FLS centred at $\alpha(2000)
=17:17:00.00$, $\delta(2000)=59:45:00.000$ was also observed at 1.4~GHz by 
Morganti et al. (2004) with the Westerbork Synthesis Radio Telescope (WSRT), 
reaching an rms noise of $\sim 8.5\mu$Jy beam$^{-1}$ in the central region, 
figure which rises to $\sim 20\mu$Jy beam$^{-1}$ at $\sim$30 arcmin 
from the centre. 1048 sources are found with peak flux 
densities brighter 
than $5\times$ the above limits, an average positional accuracy of $\sim 1.5$ 
arcsec and a beam resolution of $\sim$14 arcsec. 
These objects are represented by the (green) triangles 
in Figure~\ref{fig:sky}. We note that Morganti et al. (2004) find a 
systematic offset between their radio positions as compared to those of Condon 
et al. (2003) of about 1 arcsec. 

\begin{figure}
\includegraphics[width=8.0cm]{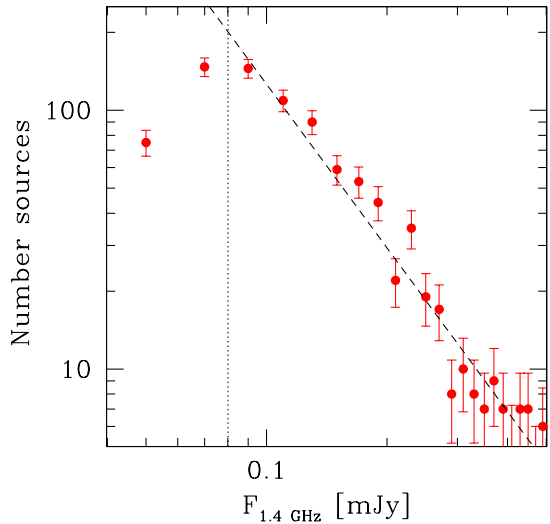}
\caption{1.4 GHz differential number counts for the Morganti et al. (2004) 
sample in $\Delta F_{1.4 \rm GHz}=0.02$~mJy bin widths. 
The vertical dotted line represents the estimated limit for 
completeness, while the dashed line is the best-fit to the data for fluxes 
$F_{1.4\rm GHz}\ge 80 \mu$Jy.
\label{fig:morganti}}
\end{figure}

The differential number counts $N(F)\equiv \Delta N/\Delta F$ as a function 
of 1.4~GHz 
integrated flux density for the Morganti et al. (2004) sample is plotted in 
Figure~\ref{fig:morganti} for a bin width $\Delta F_{1.4 \rm GHz}=0.02$~mJy. 
The data shows that the sample includes sources 
as faint as $\sim 40 \mu$Jy, with a $\sim 5\sigma$ level 
for completeness reached for fluxes $F_{1.4 \rm GHz}\simgt 80 \mu$Jy.
The best fit to the data beyond the limit for completeness is provided by the 
expression $N(F_{1.4\rm GHz})=A F_{1.4\rm GHz}^{-\gamma}$, with $A=1.0^{+0.2}_{-0.1}$ and 
$\gamma=2.1^{+0.1}_{-0.1}$ and is represented in 
Figure~\ref{fig:morganti} by the dashed line. 
The somewhat shallow slope of the counts when compared with other recent 
studies of faint radio sources (e.g. Hopkins et al. 2003; Seymour, McHardy \& Gunn 2004) 
is probably due to the non-uniform coverage of the surveyed area, also visible in Figure~2. 
However, this is not a cause for concern in the following analysis which only marginally 
relies on sample completeness.

\begin{figure}
\includegraphics[width=8.0cm]{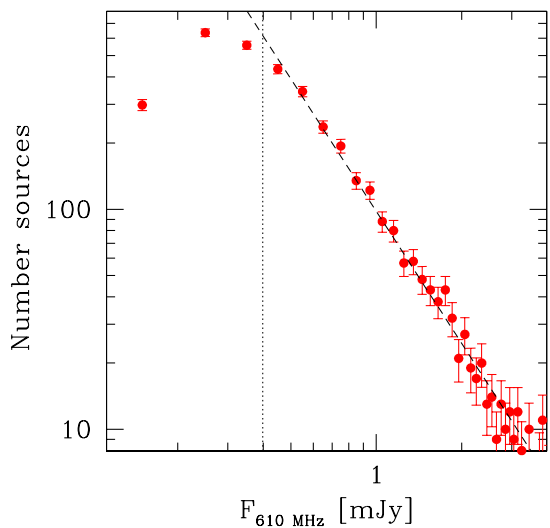}
\caption{610 MHz differential counts from the GMRT Survey (Garn et al. 2007) 
in $\Delta F_{610 \rm MHz}=0.1$~mJy bin widths. 
The vertical dotted line represents the limit for completeness, while 
the dashed line is the best-fit to the data for $F_{610\rm MHz}\ge 0.4$~mJy.
\label{fig:GMRT}}
\end{figure}

Garn et al. (2007) provide observations at 610~MHz of the First Look Survey 
field taken with the Giant Metrewave Radio Telescope (GMRT). This radio 
survey  covers a total area of $\sim 4$~deg$^2$, with a resolution of 
5.8$\times$4.7 arcsec$^2$. The rms noise $\sigma_F$ at the centre of 
the pointing is between 27 and 
30~$\mu$Jy beam$^{-1}$, figures which rise to $\sim 150\mu$Jy beam$^{-1}$ 
towards the edges. The GMRT catalogue contains 3944 sources with peak 
brightness greater than $5\sigma_F$. The sky distribution of these sources is 
presented in Figure~\ref{fig:sky} by the (blue) filled squares.

The differential number counts as a function of 610~MHz 
integrated flux density for the Garn et al. (2007) sample is plotted in 
Figure~\ref{fig:GMRT} for a bin width $\Delta F_{610 \rm MHz}=0.1$~mJy. 
The sample includes sources 
brighter than $\sim 0.1$~mJy, with a $\sim 5\sigma$ level 
for completeness reached for fluxes $F_{610 \rm MHz}\simgt 400 \mu$Jy.
The best fit to the complete part of the dataset only,
represented in Figure~\ref{fig:GMRT} by the dashed line,
is provided by the expression $N'(F_{610\rm MHz})=A' F_{610\rm MHz}^{-\gamma'}$, 
with $A'=98\pm 11$ and $\gamma'=2.0^{+0.1}_{-0.2}$.

\section{The Matched Sample}
Radio counterparts to optically obscured {\it Spitzer} sources were 
searched by cross-correlating the catalogue obtained by Magliocchetti et al. 
(2007) and described in \S 2.1, with those provided by Condon et al. 
(2003), Morganti et al. (2004) and Garn et al. (2007).

Given the large resolution beam associated to the Morganti et al. (2004) 
observations, 
in order to provide a self-consistent searching radius at both 1.4~GHz and 
610~MHz, as a first step we have decided to consider as true radio 
counterparts to 
24$\mu$m-selected, $R>25.5$ sources radio objects whose positions 
differed from those of MIPS galaxies by less than 10$^{\prime\prime}$. 
53 matches out of 510 sources (corresponding to $10.4$\% of the 
original {\it Spitzer} sample) were found in the case of the 610~MHz GMRT 
Survey over the 
whole FLS-IRAC area. On the smaller region covered by the Morganti et al. 
(2004) observations, we instead find 33 matches out of 150 optically obscured 
{\it Spitzer} sources which occupy the same portion of the sky. This 
corresponds to $\sim 22$\% of the original sample.

The chances for contamination both in the case of 1.4~GHz 
and for 610~MHz observations have been estimated by 
shifting in both RA and Dec the {\it Spitzer} sample with respect to the radio ones. 
This was repeated ten times for different shifting amounts which ranged between 1 and 5 
arcminutes. The resulting values for chance coincidences were then averaged. 
By doing this, for the Morganti et al. (2004) dataset we find 2 spurious matches, while 
in the GMRT case this figure rises to 7 objects. These values are somewhat lower than 
what statistically predicted by considering the surface density of both radio and MIPS 
sources: for the 610~MHz sample we expect $\sim 12$ spurious matches, while 
for the Morganti et al. (2004) sample this figure is $\sim 4$. One possible explanation 
for such a discrepancy could be found in the non-uniform coverage of both the Garn et al. 
(2007) and Morganti et al. (2004) surveys (\S2.2). As a matter of fact, 
there are only 6 sources with an offset between 610~MHz and 24$\mu$m 
positions greater than 6 arcsec, size of the GMRT beam and radius at which the chances for 
contamination drop to $\simlt 4$~\%. They are: J171948.6+585133, J171810.7+591639, 
J172317.5+591109, J171628.4+601342, J172122.3+600605, J172042.6+590930. These objects have 
been checked by eye on the 1.4~GHz maps provided online by Condon et al. 
(2003; url: $http://www.cv.nrao.edu/sirtf/$) and in all cases they 
were found to likely be real associations.

\begin{figure}
\includegraphics[width=8.0cm]{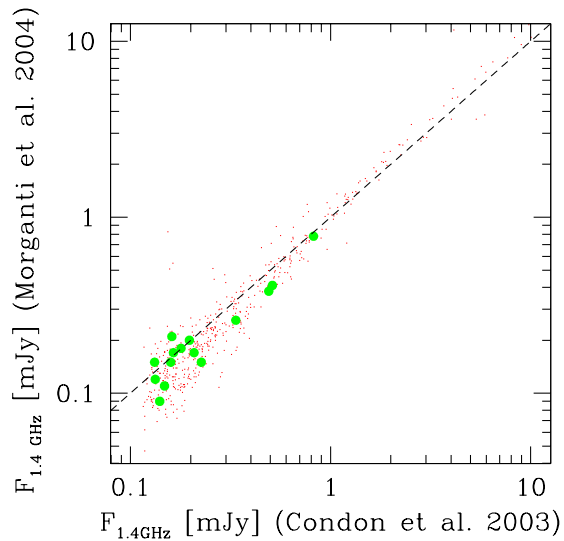}
\caption{1.4~GHz integrated flux densities for the optically obscured 
{\it Spitzer} sources presented in this work as reported by Morganti et al. (2004; y-axis) 
and Condon et al. (2003; x-axis) in the portion of sky where these two surveys 
overlap (green circles). The small dots represent the Condon et al. (2003) 
and Morganti et al. (2004) flux measurements for {\it all} radio sources which appear 
in both catalogues.
\label{fig:morgantivscondon}}
\end{figure}

Finally, we also used the Condon et al. (2003) catalogue provided online and 
searched for 1.4~GHz 
counterparts to optically obscured {\it Spitzer} galaxies within the same 
matching radius of 10$^{\prime\prime}$ as indicated above over the whole 
FLS-IRAC region. The number of sources with measured radio fluxes from 
the Condon et al. (2003) online catalogue 
is 70, corresponding to 13.7\% of the original sample. 
Also in this case we have visually checked those radio-to-24$\mu$m associations 
having distances between 6$^{\prime\prime}$ and 10$^{\prime\prime}$ and found them to 
likely be real ones. These objects are: J171529.9+593448, J171656.7+594103, J172317.5+591109, 
J171511.5+585741, J172050.3+590638, J172005.0+592430 and J172042.6+590930.

\begin{figure*}
\includegraphics[width=8.0cm]{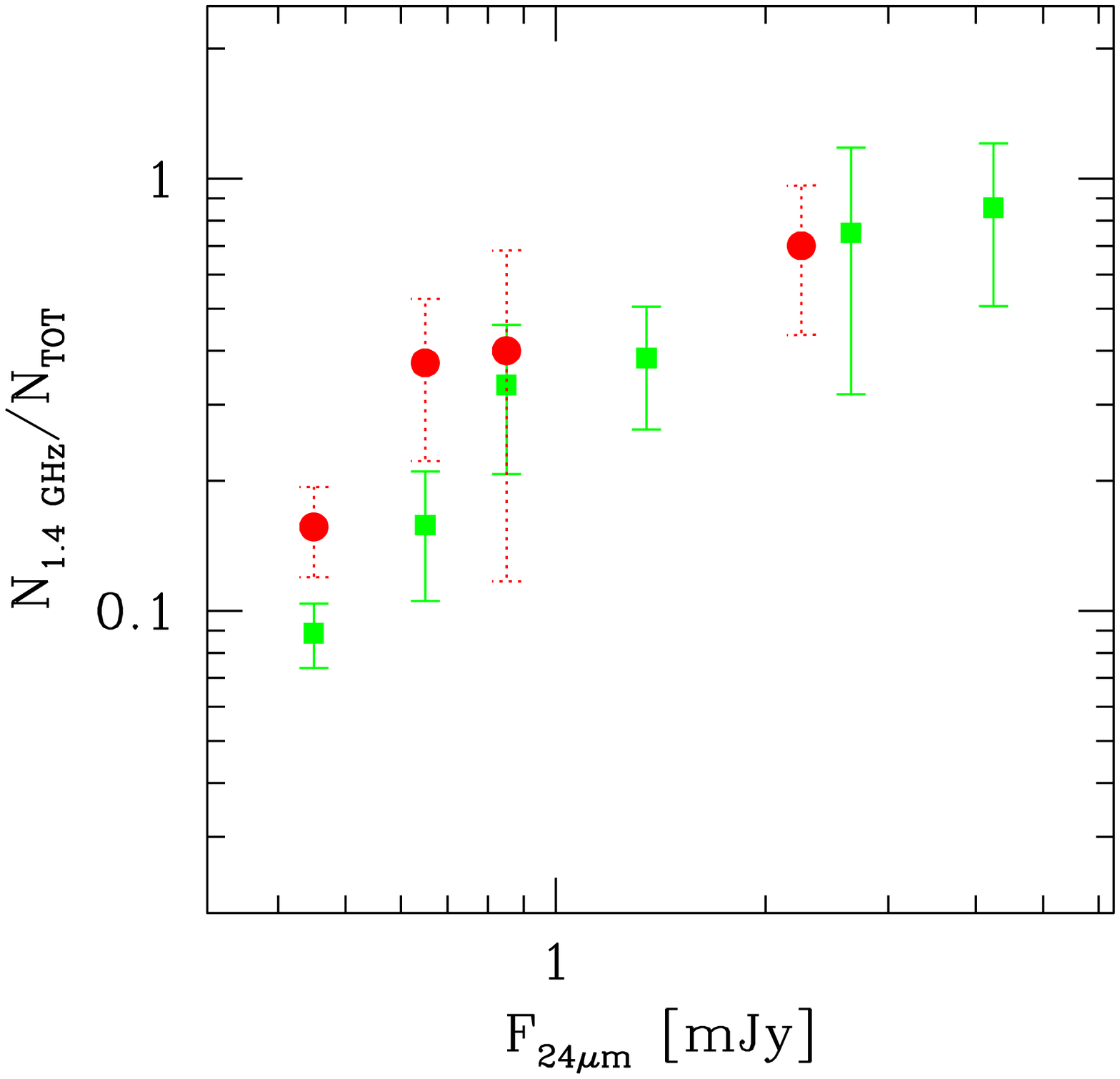}
\includegraphics[width=8.0cm]{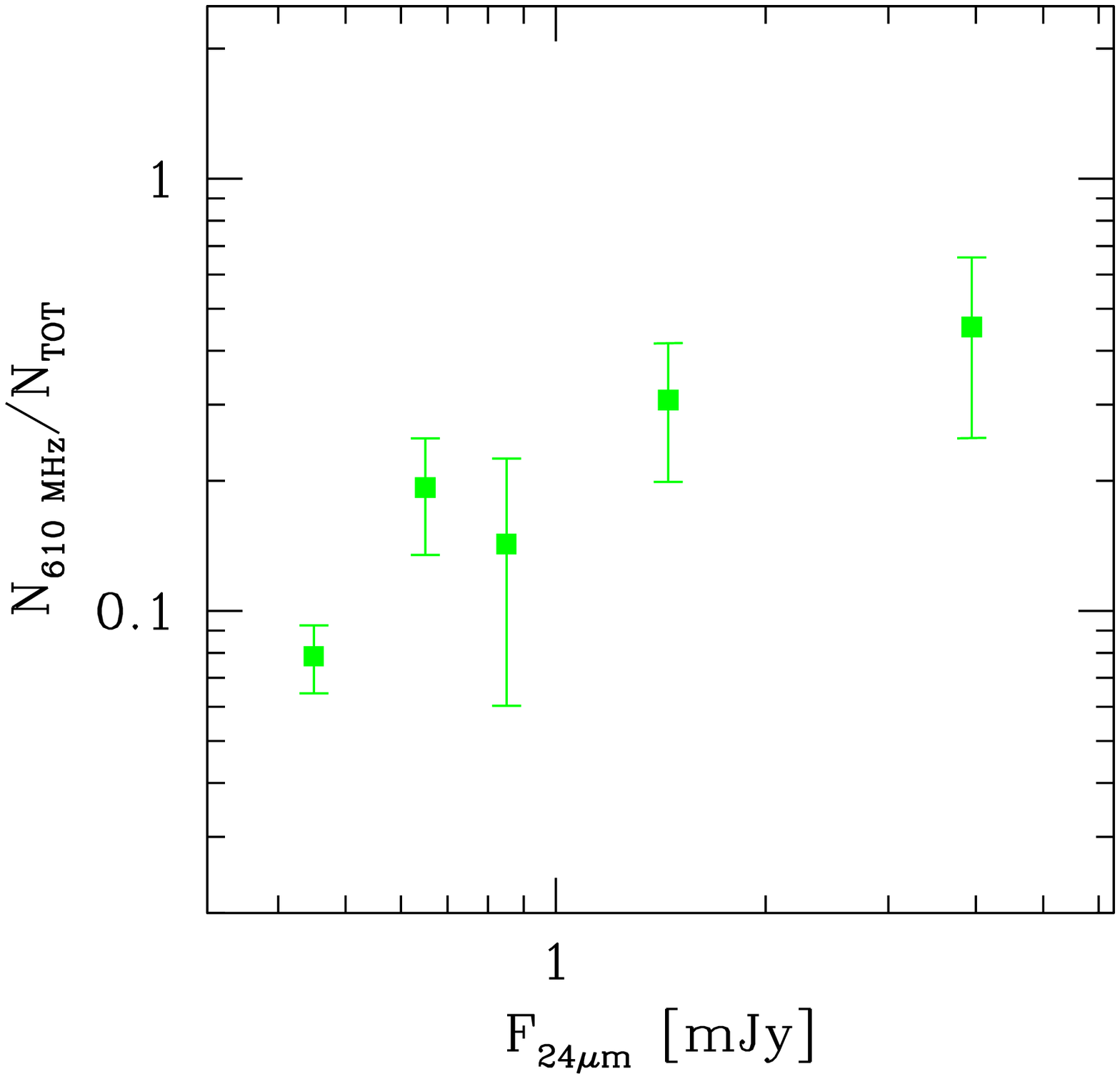}
\caption{Left-hand panel: fractional number of 1.4 GHz counterparts to 
obscured {\it Spitzer} sources as a function of 24$\mu$m flux. The (green) 
squares are for the Condon et al. (2003) sample, while the 
(red) dots are for the Morganti et al. (2004) dataset. Error-bars 
represent the 1$\sigma$ Poisson uncertainties on the number of sources. 
Right-hand panel: same as before but for the GMRT survey (Garn 
et al. 2007).
\label{fig:radioid}}
\end{figure*}

Fifteen sources belonging to the original $R\simgt 25.5$, {\it Spitzer} 
sample have 1.4~GHz counterparts 
both in the Condon et al. (2003) and Morganti et al. (2004) catalogues. 
A comparison of their integrated
radio fluxes (performed in Figure~\ref{fig:morgantivscondon}) as taken 
from these two datasets, shows that -- except for a few cases -- 
the Morganti et al. (2004) fluxes are systematically lower than 
those measured by Condon et al. (2003). 
We do not know the reason for this discrepancy which, given the larger beam 
of the WSRT, if anything should have gone the other way round 
(i.e. Condon fluxes smaller than the Morganti et al. ones). 
A possible cause could be found in calibration problems which may have affected 
flux measurements for faint sources in one or another survey. The same trend for fluxes as 
coming from the Morganti et al. (2004) catalogue to be systematically lower than those 
measured by Condon et al. (2003) is in fact observed for {\it all} radio sources fainter than 
$\sim$1~mJy which are present in both datasets (520 to a distance of 6 arcsec; see 
Figure~\ref{fig:morgantivscondon}).

For the 15 {\it Spitzer} sources belonging to our sample with double 1.4~GHz measurements, 
we then decided to adopt as 'true' fluxes those given by Condon et al. (2003) as it is in this sample where we find the overwhelming majority of 1.4~GHz 
counterparts to our {\it Spitzer} sources. 
We note tha most of the following analysis and discussion, especially that devoted 
to the radio spectral indices in \S6, is only very marginally affected by the eventual 
inclusion of the information coming from the Morganti et al. (2004) dataset.

Out of the remaining 18 optically faint {\it Spitzer} objects with 
counterparts in 
the Morganti et al (2004) catalogue, 16 of them have radio fluxes smaller than 
$\sim 130 \mu$Jy, consistent with the fact that we do not find them in the 
Condon et al. (2003) dataset. However, two objects, quite bright in the 
Morganti et al. (2004) radio maps did not seem to have been included in the 
Condon et al. (2003) catalogue. These are J171427.8+592828 
($F_{1.4\rm GHz}=0.83$~mJy) and 
J171527.5+593139 ($F_{1.4\rm GHz}=0.39$~mJy). Direct visual 
investigations of the radio maps provided online by Condon et al. (2003) 
show that in the first case the source has been missed in the matching 
procedure with the Condon et al. (2003) catalogue 
as presenting a distance between radio and {\it Spitzer} counterpart 
greater than the adopted 10$^{\prime\prime}$ matching radius. 
The association however looks real, as the 
24$\mu$m emission comes from a lateral blob of the radio source 
(see the postage map in the middle of the bottom row in 
Figure~\ref{fig:bright}). Also J171527.5+593139 exists in the 
Condon et al. (2003) radio maps, even though they 
measure a 1.4~GHz radio flux which is faint enough to have it excluded 
from their $>5\sigma_F$ catalogue. We therefore also take this 
radio-to-{\it Spitzer} 
association as real, even though we warn the reader on the reliability 
of its 1.4~GHz flux value. 

Finally, with the help of the higher positional accuracies and smaller 
beam resolutions of the Condon and IRAC surveys with respect to those of 
the 24$\mu$m MIPS channel (see \S2.2 and \S4), we find that there were three objects in the 
radio-{\it Spitzer} catalogue which were erroneously split into a number 
of different sources. These are J171628.4+601342, which the 
deconvolution technique adopted for MIPS sources had divided into three 
different objects, and J172018.1+592902 and J172353.2+601354, which both appear in 
the original {\it Spitzer} catalogue as two different objects. However, all of them only had 
one identification at both 1.4~GHz and in the IRAC channels. In these cases, 
the 24$\mu$m flux associated to the single {\it Spitzer} source was simply 
obtained by adding the 24$\mu$m fluxes of its sub-components.

The result of the above 'cleaning procedures' leaves 67 radio identified 
optically faint {\it Spitzer} sources from the Condon et al. (2003) catalogue, 
17 from the Morganti et al. (2004) catalogue and 52 from the Garn et al. (2007) dataset. 
To this last figure we have also added 5 more sources (i.e. J171054.4+594426, 
J172005.0+592430, J172103.6+585052, J171427.8+592828, J172217.4+601003) 
which had bright ($F_{1.4\rm GHz}\simgt 0.8$~mJy) 
1.4~GHz counterparts but no measured 610~MHz fluxes from the automatic 
matching procedure. Visual investigations of the 
radio maps have in fact shown that -- in all but one of these five 1.4~GHz-identified 
sources -- the radio emission is extended and the 24$\mu$m signal 
stems from regions which correspond to secondary peaks in the radio emission. 
This explains the 610~MHz-to-24$\mu$m association to be found at distances 
slightly larger than 10$^{\prime\prime}$. Images for these four cases 
are presented in Figure 15 (top-left, top-right, centre and bottom-centre), 
while all the five sources are furtherly discussed in the Appendix.

The final number of optically faint {\it Spitzer} sources with a radio 
counterpart either at 1.4~GHz or at 610~MHz or at both frequencies is 96. 
This is the sample which we will use for our studies throughout the paper.
All these sources were visually checked in the Condon et al. (2003) radio 
maps to assess the likelihood of the associations, investigate  
peculiarities in the radio-to-24$\mu$m emission and look for interesting 
features in their radio morphologies. 
Except for one dubious case (namely J172217.4+601003), 
all the other radio-to-24$\mu$m associations look real. 
 
The mid-IR and radio properties for these sources are summarized in Table~1.
The Appendix instead provides images and a more detailed description for 
the most peculiar objects. 
Except for J171143.9+600741 and J172005.0+592430 which are two 
very extended and bright triple sources, in all the other cases of 
(more dubious) multi-components, the radio fluxes reported in Table~1 and 
used throughout this work simply correspond to the flux of the radio source 
which is closest to the centre of 24$\mu$m emission, i.e. no collapsing 
technique has been applied to these objects. However, images and a 
detailed description for all of them are given in the Appendix.
In passing, we note that in most of the brightest radio sources (i.e. 
brighter than $F_{1.4\rm GHz}\sim 0.8$~mJy, see Figure 15), the mid-IR emitting region 
is associated to secondary peaks of radio emission such as 
 radio lobes and jets or distant star-forming regions 
rather than steming from the site of primary radio emission, 
most likely corresponding to the position of the accreting 
black hole. 
If these associations are indeed real, they surely deserve 
a deeper investigation which will be the subject of a forthcoming paper.

The fraction of radio-identified/optically obscured {\it Spitzer} sources 
as a function of 24$\mu$m flux is shown in 
Figure~\ref{fig:radioid}, left-hand panel for what concerns 1.4~GHz 
observations and right-hand panel for 610~MHz ones. In the left-hand panel, 
(green) squares correspond to the Condon et al. (2003) data, while the (red) 
dots are for the Morganti et al. (2004) measurements. Almost all the sources 
with 24$\mu$m fluxes brighter than $\sim 2$~mJy have been identified at 
1.4~GHz. In more detail, we have that only one object brighter than the above 
limit does not have a 1.4~GHz counterpart either in the Morganti et al. (2004) 
or in the Condon et al. (2003) catalogues (see also Figure~\ref{fig:frvsf24}). 
However, such a high completeness level for radio identifications 
quickly drops as one moves to fainter 24$\mu$m fluxes, especially 
if the radio counterpart is searched for within radio objects brighter than 
$F_{1.4\rm GHz} \sim 0.1$~mJy as it is the case for the Condon et al. (2003) 
data. The fraction of radio-identified {\it Spitzer} sources instead remains 
quite high (on the order of 50\%) at all $F_{24\mu\rm m}\simgt 0.5$~mJy fluxes 
if one considers objects with 1.4~GHz fluxes below $\sim 0.1$~mJy 
as those probed by the Morganti et al. (2004) survey. 

The situation is quite different if one considers the case for 610~MHz 
counterparts to optically obscured {\it Spitzer} sources. In fact, due to the 
higher flux limit of the GMRT survey with respect to the 1.4~GHz ones, 
the fraction of radio identified objects 
here is much lower, less than 50\% even at the highest 24$\mu$m fluxes (see 
Figure~\ref{fig:radioid}). 
This fraction then dramatically drops to less than 20\% already 
below $F_{24\mu\rm m}\sim 1$~mJy.

\section{IRAC photometry}
As already mentioned in \S 2.1, the FLS area is covered by IRAC 
observations performed at 3.6, 4.5, 5.8 and 8$\mu$m. To look for IRAC 
counterparts to the optically faint sources presented in \S 2.1, we have then 
relied on the band-merged catalogue provided by Lacy et al. (2005) and 
cross-correlated 24$\mu$m and IRAC positions by requiring a matching radius 
smaller than 2 arcsec.

\begin{figure}
\includegraphics[width=8.0cm]{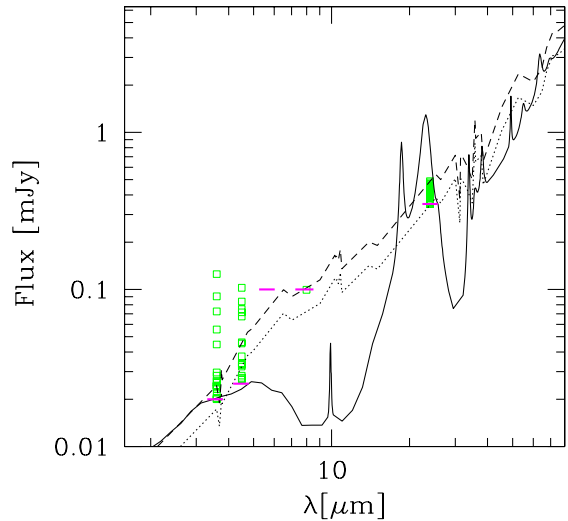}
\caption{IRAC fluxes for $R>25.5$ {\it Spitzer} sources 
fainter than $F_{24\mu \rm m}=0.5$~mJy and belonging to our 
radio-identified sample (green squares). The solid line represents 
an Arp220-like SED redshifted at $z=2$, while the dashed and dotted lines 
are for two Mkn231-like SEDs at $z=2$ with different 24$\mu$m normalizations 
(0.5 mJy for the upper curve, 0.35 mJy for the lower one). 
The horizontal (magenta) dashes indicate the 5$\sigma$ flux limits of the IRAC and MIPS 
data on the FLS area.
\label{fig:SED} }  
\end{figure}

323 sources out of 510 (corresponding to $\sim$63\% of the original sample) 
have measured IRAC 
fluxes, at least in one of the two deeper 3.6 or 4.5 $\mu$m bands 
(we remind the reader that the 5$\sigma$ flux limits in the four IRAC channels are 
20~$\mu$Jy, 25~$\mu$Jy, 100~$\mu$Jy, 100~$\mu$Jy, respectively at 3.6~$\mu$m, 
4.5~$\mu$m, 5.8~$\mu$m and 8~$\mu$m, see Figure \ref{fig:SED}). 
If one only concentrates on those 
sources which belong to our radio-identified catalogue, the number of 
objects with an IRAC counterpart is 64, value which corresponds to 
67\% of its parent sample. IRAC photometry for such sources is reported in 
Table~1.

IRAC fluxes constitute an important tool to help understanding the nature of 
the objects under examination. In fact, as for instance shown by 
Magliocchetti et al. (2007), a diagnostic based on the ratio between 
8$\mu$m and 24$\mu$m fluxes can discriminate between objects which are 
mainly powered by intense star-formation -- featuring a Spectral Energy Distribution 
(SED) similar to that of Arp220 -- and those dominated by an obscured AGN -- 
described by a SED similar to that of Mkn231 --, the 
dividing line between these two categories being set by a ratio 
$F_{8\mu \rm m}/F_{24\mu\rm m}\sim 0.1$, whereby starbursts are found below this value.. 
Unfortunately, due to the relatively bright flux limit of the 8$\mu$m channel 
of the FLS, such a diagnostic is only valid for objects with 24$\mu$m fluxes 
brighter than $\sim 0.5-0.6$~mJy, as AGN-powered sources fainter than this 
will be undetected at 8$\mu$m (see Figure~\ref{fig:SED}) 
and therefore will present ratios $F_{8\mu \rm m}/F_{24\mu\rm m}< 0.1$, 
compatible with those of star-forming galaxies. 


\begin{figure}
\includegraphics[width=8.0cm]{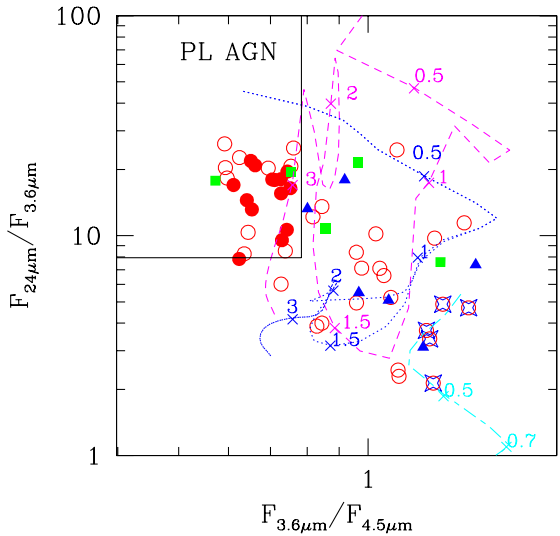}
\caption{Distribution of $F_{3.6\mu\rm m}/F_{4.5\mu\rm m}$ vs $F_{24\mu\rm m}/F_{3.6\mu\rm m}$ 
flux ratios for those 57 radio-detected, optically faint {\it Spitzer}-FLS 
sources which present an IRAC counterpart at both 3.6 $\mu$m and 4.5 $\mu$m. 
These are compared with the computed colour-colour tracks as a function of redshift 
for four SEDs: that of Arp220 (magenta/dashed line), that of M82 (lower blue/dotted line), 
that of a Mkn231-like galaxy with an added evolved stellar contribution 
(upper blue/dotted line) and that of M51 (cyan short/long dashes). 
Some reference redshift values are marked by crosses along the SEDs. Circles 
(both empty and filled) are for the class of AGN, squares for SF galaxies, while 
triangles represent unclassified objects and circles+stars low-z contaminants. 
The region marked by the box encloses those AGN which feature a power-law shape in their 
SED over the whole 3.6$\mu$m-24$\mu$m  wavelength range (PL AGN; see text for details).
\label{fig:z}}
\end{figure}

For 24$\mu$m fluxes fainter than the above limit, one can instead try to rely 
on the two deeper 3.6 and 4.5$\mu$m channels of IRAC to investigate the 
nature of these optically obscured {\it Spitzer} sources. 
In fact, in absence of an evolved stellar population which at redshift 
$z\sim 2$ would determine a 'bump' in the SED between $\sim 3\mu$m and 
$\sim 8\mu$m (see Figure~\ref{fig:SED}), 
the Spectral Energy Distribution of AGN-powered sources features a nice 
power-law with slope $\simgt 1$ all through the IRAC channels (see e.g. Stern et al. 2005). 
Optically faint {\it Spitzer} sources with $F_{24\mu \rm m}\le 0.5$~mJy have 
then been classified as AGNs if $F_{4.5\mu\rm m}/F_{3.6\mu \rm m}\ge 4.5/3.6$. 
All the other objects, with either no available IRAC photometry or 
with $F_{4.5\mu\rm m}/F_{3.6\mu \rm m}$ ratios shallower than the adopted 
value are instead to be considered as unclassified since 
an evolved stellar population could be present in both star-forming 
galaxies and AGN-dominated systems.

The addition of the two lowest wavelength IRAC channels can also help us finding low redshift 
contaminants. To this aim, the distribution of $F_{3.6\mu\rm m}/F_{4.5\mu\rm m}$ vs 
$F_{24\mu\rm m}/F_{3.6\mu\rm m}$ colours for the whole sample of optically obscured {\it Spitzer} 
sources with a counterpart at both 3.6$\mu$m and 4.5$\mu$m  
was compared to the computed colour-colour tracks as a function of redshift 
for five different SEDs which most likely represent the dominant populations selected in a 24$\mu$m 
survey: those of Arp220 and M82 (respectively very extreme and more quiescent starforming systems), 
that of the spiral galaxy M51 and those of an obscured AGN, both with and without a contribution coming 
from an evolved stellar population.  All the tracks are plotted in Figure~\ref{fig:z}, and the 
latter two are respectively indicated by the whole area enclosed by the box on the upper-left quadrant 
of Figure~\ref{fig:z} (indicating pure power-law SED AGN with mid-IR slopes $\beta>1$) and by the upper 
blue/dotted curve (example of a Mkn231-like SED with an added contribution from an evolved stellar 
population). 

The above analysis, together with the results presented in Magliocchetti et al. 2007, 
allows to identify the following categories for 24$\mu$m-selected sources without an optical 
counterpart:\\
A) Parent MIPS-IRAC catalogue (510 sources):\\
- AGN0; $F_{24\mu \rm m}\ge 0.6$;  $F_{8\mu\rm m}/F_{24\mu \rm m}\ge 0.1$; 
104 objects.\\
- SB0; $F_{24\mu \rm m}\ge 0.6$;  $F_{8\mu\rm m}/F_{24\mu \rm m}< 0.1$; 
31 objects.\\
- AGN1; $F_{24\mu \rm m}< 0.6$;  $F_{4.5\mu\rm m}/F_{3.6\mu \rm m}\ge 4.5/3.6$;
 117 objects.\\
- Unclassified; $F_{24\mu \rm m}< 0.6$;  $F_{4.5\mu\rm m}/F_{3.6\mu \rm m}
< 4.5/3.6$ or no IRAC information; 258 objects.\\
B) Radio-identified catalogue (96 sources):\\
- AGN0; $F_{24\mu \rm m}\ge 0.6$;  $F_{8\mu\rm m}/F_{24\mu \rm m}\ge 0.1$; 
29 objects.\\
- SB0; $F_{24\mu \rm m}\ge 0.6$;  $F_{8\mu\rm m}/F_{24\mu \rm m}< 0.1$; 13 
objects.\\
- AGN1; $F_{24\mu \rm m}< 0.6$;  $F_{4.5\mu\rm m}/F_{3.6\mu \rm m}\ge 
4.5/3.6$; 16 objects.\\
- Unclassified; $F_{24\mu \rm m}< 0.6$;  $F_{4.5\mu\rm m}/F_{3.6\mu \rm m}
<4.5/3.6 $ or no IRAC information; 33 objects.\\
The distribution of $F_{3.6\mu\rm m}/F_{4.5\mu\rm m}$ vs 
$F_{24\mu\rm m}/F_{3.6\mu\rm m}$ colours for the subsample of radio-identified sources is 
presented in Figure~\ref{fig:z}. Open circles are for the subclass of AGN0, while filled ones for AGN1.
We note that in the subsample of radio-identified sources there are five low-z contaminants 
(i.e. $z\simlt 0.5$ galaxies most likely of the M51-type, hereafter denoted as local interlopers or LI). 
These sources, unless explicitly 
stated, will never be included in the following analysis and discussions. We also note that there are 
three dubious cases of $z\simlt 1$ sources, which were also classified as AGN on the basis of their 
$F_{8\mu\rm m}/F_{24\mu \rm m}$ ratios but which could also possibly be 
M82-like galaxies. 
The various classes for the radio-identified sources (i.e. whether AGN0, AGN1, SF, UNCL or LI) 
are reported in Table~1. The three dubious AGN cases are denoted with an 'AGN0?'

IRAC photometry can also be used to provide rough redshift estimates.
In fact, Magliocchetti et al. (2007) could only perform a very broad distinction 
between local {\it Spitzer} sources and high redshift (i.e. $z\simgt 1.6$) 
ones based on comparisons between 24$\mu$m-vs-R colours and computed colour-colour 
evolution tracks obtained for a number of representative dusty SEDs such as 
Arp220 and Mkn231. Even though an indirect confirmation that at least the majority 
of such sources were indeed at high redshifts came from comparisons of 
the observed number counts with predictions from models for galaxy formation 
and evolution (e.g. Granato et al. 2004), the approach adopted by 
Magliocchetti et al. (2007) could only be considered as a zero-order one. 
As already noticed in the previous discussion, chances for contamination in the high-z sample are 
in fact non-negligible, especially in the case of obscured AGN which present power-law SEDs in 
the mid-IR region and are therefore more difficult to locate in redshift.

By making use once again of the colour-colour evolutionary tracks indicated in Figure~\ref{fig:z}, 
we can attempt at assigning broad redshift groups to all those sources belonging to the 
Magliocchetti et al. (2007) sample whose SEDs present a contribution from an evolved stellar 
population in the two 3.6$\mu$m and 4.5$\mu$m bands.
In fact, as Figure~\ref{fig:z} shows, this method is quite robust, 
as - for a chosen (wide) redshift interval -- all the adopted SEDs present similar values for 
the colours in the [3.6-4.5]$\mu$m range.
The results of this analysis can be summarized as follows: \\
z-type 1 ($0\simlt z\simlt 1$): 52 (total sample), 11 (radio-id sample);\\
z-type 2 ($1\simlt z\simlt 1.5$): 36 (total sample), 7 (radio-id sample);\\
z-type 3 ($1.5\simlt z\simlt 3$): 54 (total sample), 13 (radio-id sample);\\
z-type 4 ('pure' $\beta>1$ PL AGN SED; no redshift assignment is possible in these cases): 
111 (total sample), 27 (radio-id sample),\\ 
and show, as expected, that the majority of optically obscured {\it Spitzer} sources 
resides at redshifts $\simgt 1$ ($\sim 65$\% of the sub-class of galaxies 
with assigned redshift), 
both for the whole population of optically faint {\it Spitzer}-selected 
sources and for the radio-identified ones). Perhaps not surprising, 
we have that almost all $z\simlt 1.5$ sources in our sample are AGN which, 
because of their SEDs were the most difficult category to be correctly set at 
high redshifts. On the other hand, this implies that our result 
on the fraction of $z\simgt 1$ sources  
is most likely a lower limit: indeed the overwhelming 
majority of those objects classified as starforming galaxies do not have measured 
fluxes in the IRAC channels, probably indicating that they
still haven't had time to evolve a stellar population.  
However, these galaxies are the most likely to be correctly 
assigned to the $\sim[1.6-2.7]$ redshift range because of the characteristic 
rest-frame 8$\mu$m PAH region of their spectra (see Figure~6). The inclusion of these 
sources is then definitely expected to boost the fraction 
of $z\simgt 1$ sources as found in this Section by a sensible amount. 

It is also interesting to notice that 
a significant portion of those objects classified as AGN-dominated present 
an evolved stellar population (circles which occupy the lower-right quadrant of 
Figure~\ref{fig:z}): this happens for $\sim 50$\% 
of the radio identified objects despite the bias introduced in 
our $F_{24\mu\rm m}\simlt 0.6$~mJy classification criteria which identify as AGN only those 
sources with $\beta>1$. This lower limit provides an indirect evidence 
that the AGN-dominated phase in a massive galaxy takes place once 
the major episode for star-formation has ended, leaving the source with a 
population of old stars (see e.g. Granato et al. 2004).

Values for the z-types for the sources in our sample are given in Table~1.
A z-type 5 implies that no redshift estimate was possible in that case.
The above results on both classification and redshift determination will be 
handy in the next sections when we will investigate the nature of radio 
emission in these sources.


\section{Radio Properties}
\subsection{Number Counts}
\begin{figure*}
\includegraphics[width=8.0cm]{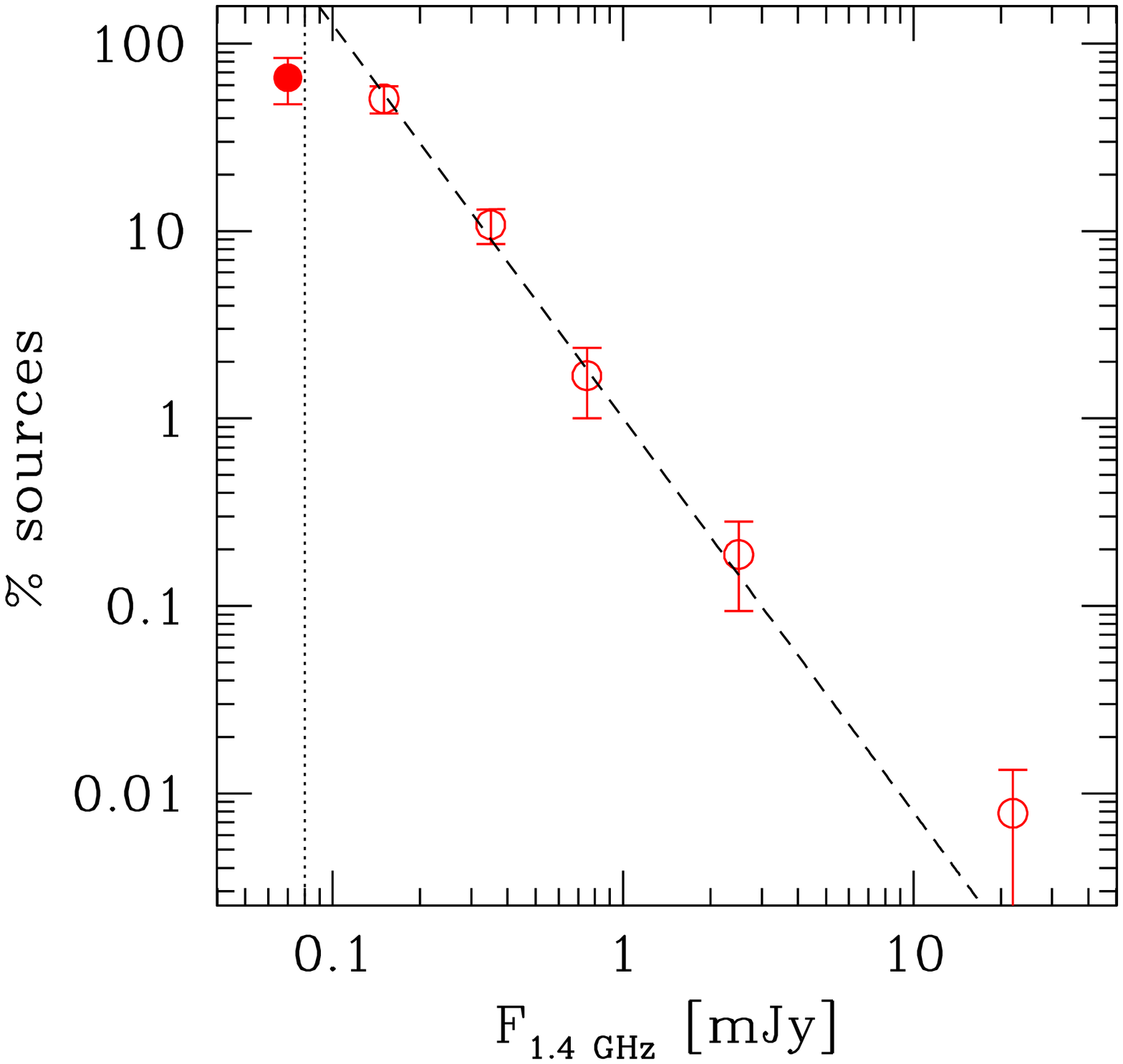}
\includegraphics[width=8.0cm]{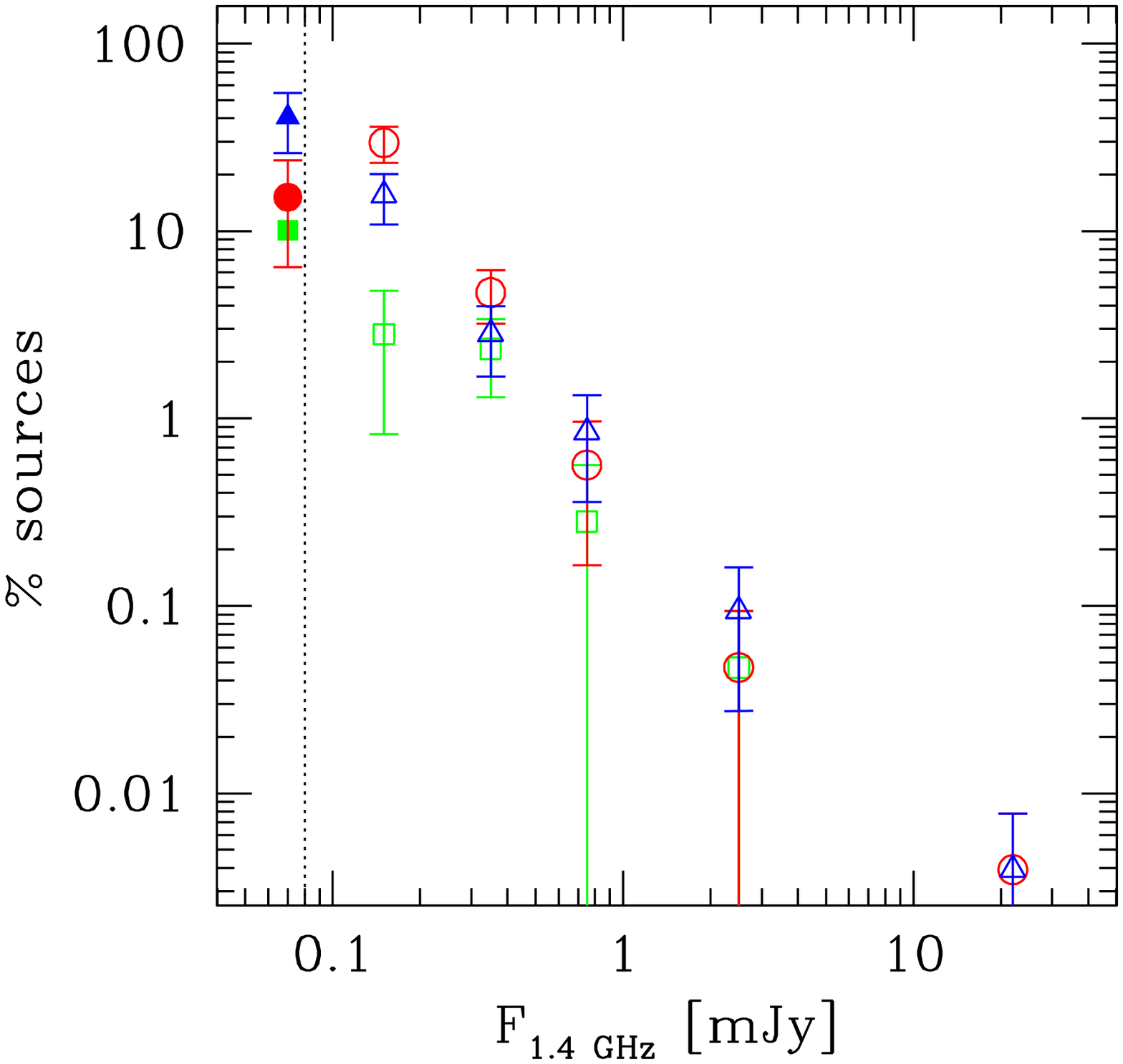}
\caption{Fractional number of obscured {\it Spitzer} sources with a radio 
counterpart vs 1.4 GHz flux. On the left-hand panel we report the total 
counts, while the plot on the right-hand side shows the counts as obtained 
for different types of objects: (red) circles are for AGN, (green) squares 
for star-forming 
galaxies and (blue) triangles for un-classified objects. Open symbols 
correspond  to the Condon et al. (2003) radio measurements, filled ones to 
the Morganti et al. (2004) dataset. The dotted lines indicate the limit for 
1.4 GHz radio completeness, while the dashed line in the left-hand plot is the best 
fit to the data beyond such a limit. 
\label{fig:counts1.4}}
\end{figure*}

\begin{figure*}
\includegraphics[width=8.0cm]{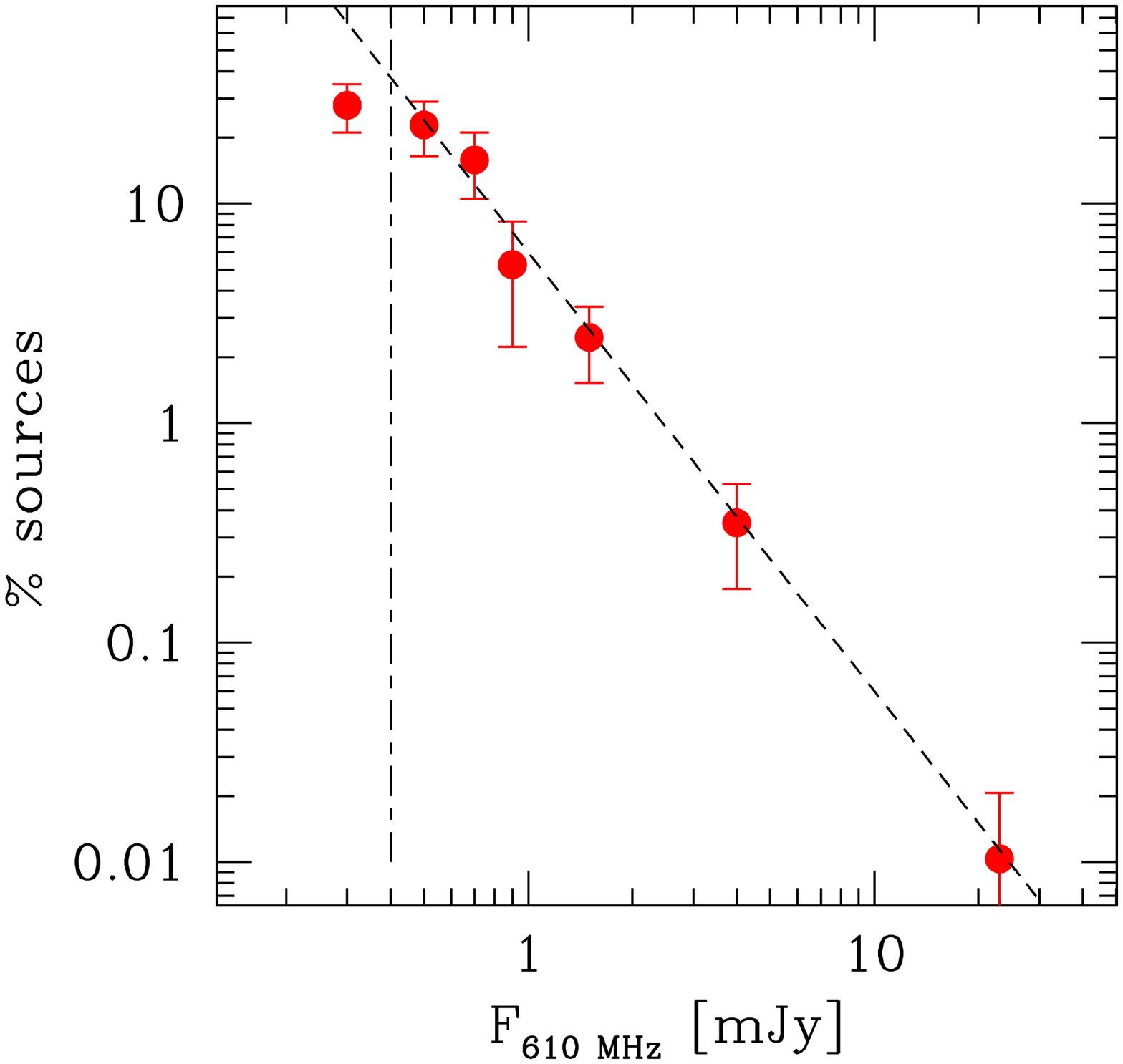}
\includegraphics[width=8.0cm]{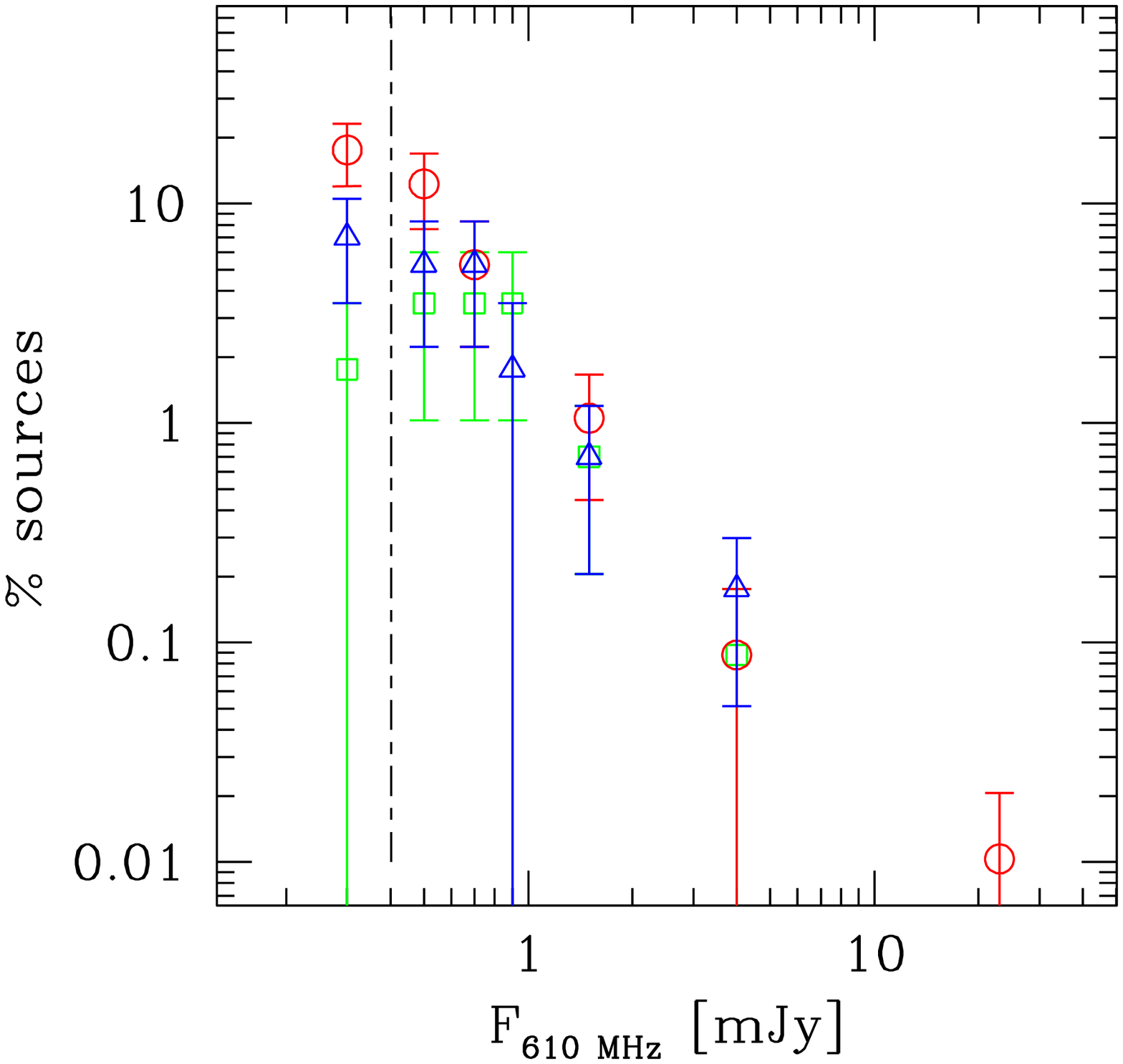}
\caption{Fractional number of obscured {\it Spitzer} sources with a radio 
counterpart vs 610 MHz flux. On the left-hand panel we report the total 
counts, while the plot on the right-hand side shows the counts as obtained 
for different types of objects: (red) circles are for AGN, (green) squares 
for star-forming galaxies and (blue) triangles for un-classified objects.
The long-short dashed lines indicate the limit for GMRT
radio completeness, while the dashed line in the left-hand plot is the best 
fit to the data beyond such a limit.
\label{fig:counts610}}
\end{figure*}

Figures~\ref{fig:counts1.4} and \ref{fig:counts610} present the differential 
radio number counts for those optically faint {\it Spitzer} sources 
respectively identified at 1.4~GHz and 610~MHz. In this first case, the chosen 
bin width is $\Delta F_{1.4\rm GHz}=0.1$~mJy, while for 610~MHz observations 
we adopted $\Delta F_{610\rm MHz}=0.2$~mJy. In both plots, the number of 
sources per flux interval has been normalized to the total number of radio 
identifications at that frequency and the data coming from the Morganti et 
al. (2004) catalogue have also been corrected so to take into account the 
smaller area covered by the survey. The left-hand panels of both 
Figures~\ref{fig:counts1.4} and \ref{fig:counts610} show the total 
number counts, while the right-hand ones illustrate the counts for the 
different categories, i.e. whether the source is to be considered an AGN, a 
star-forming galaxy or is unclassified. The different classifications, based 
on mid-IR photometry, come from \S4 and the class of AGN includes all the 
sources denoted as both AGN0 and AGN1. 

Two main features can be noticed from these plots. The first 
one is that the radio counts, both at 1.4~GHz and at 610~MHz of sources 
classified as star-forming galaxies and AGN are similar at all 
radio fluxes, especially if one adds together the $\Delta N/\Delta F$ contributions 
coming from 
star-forming galaxies and unclassified objects. Indeed, as already observed in 
\S2.1, IRAC observations 
which are deeper than those taken on the FLS show that the population 
which dominates the faint 24$\mu$m counts -- in our case constituting 
the bulk of what we call the class of unidentified objects -- is made 
of star-forming galaxies (see e.g. Brand et al. 2006; Weedman et al. 2006; 
Treister et al. 2006; Magliocchetti et al. 
2007a). This similarity between the radio counts of star-forming objects and 
AGN at both radio frequencies is in striking disagreement with the results 
coming from mid-IR photometry which show a great discrepancy e.g. in the 
differential 24$\mu$m counts of these two classes of sources, whereby  
candidate AGNs constitute the preponderant population at bright fluxes, while 
-- as discussed above -- star-forming objects dominate the faint counts (see 
Figure~4 of Magliocchetti et al. 2007). \\
Although one cannot exclude some kind of cosmic conspiracy, such a similarity 
between the radio counts of star-forming galaxies and AGNs suggests that the 
origin(s) of the radio signal at the chosen frequencies is the same for both 
classes of objects. We will investigate this issue further in the next 
sections.

Some more information on the nature of radio emission in these sources can be 
provided by comparing the total (i.e. obtained for all the radio-identified objects, 
independent on their class) differential number counts at the different radio frequencies. 
A $\chi^2$-analysis performed on the 1.4~GHz counts for fluxes above 
the inferred completeness level (\S2.2), shows that the best description 
to the data is provided by a functional form of the kind $N(F)=
A_S F_{1.4\rm GHz}^{-\gamma_S}$, with $A_S=1.0^{+0.2}_{-0.3}$ and 
$\gamma_S=2.1^{+0.1}_{-0.2}$. A similar approach applied to the 
610~MHz-detected sample instead brings $A_S^\prime=6^{+1}_{-2}$ and 
$\gamma_S^\prime=2.0^{+0.5}_{-0.6}$. 

\begin{figure*}
\includegraphics[width=8.0cm]{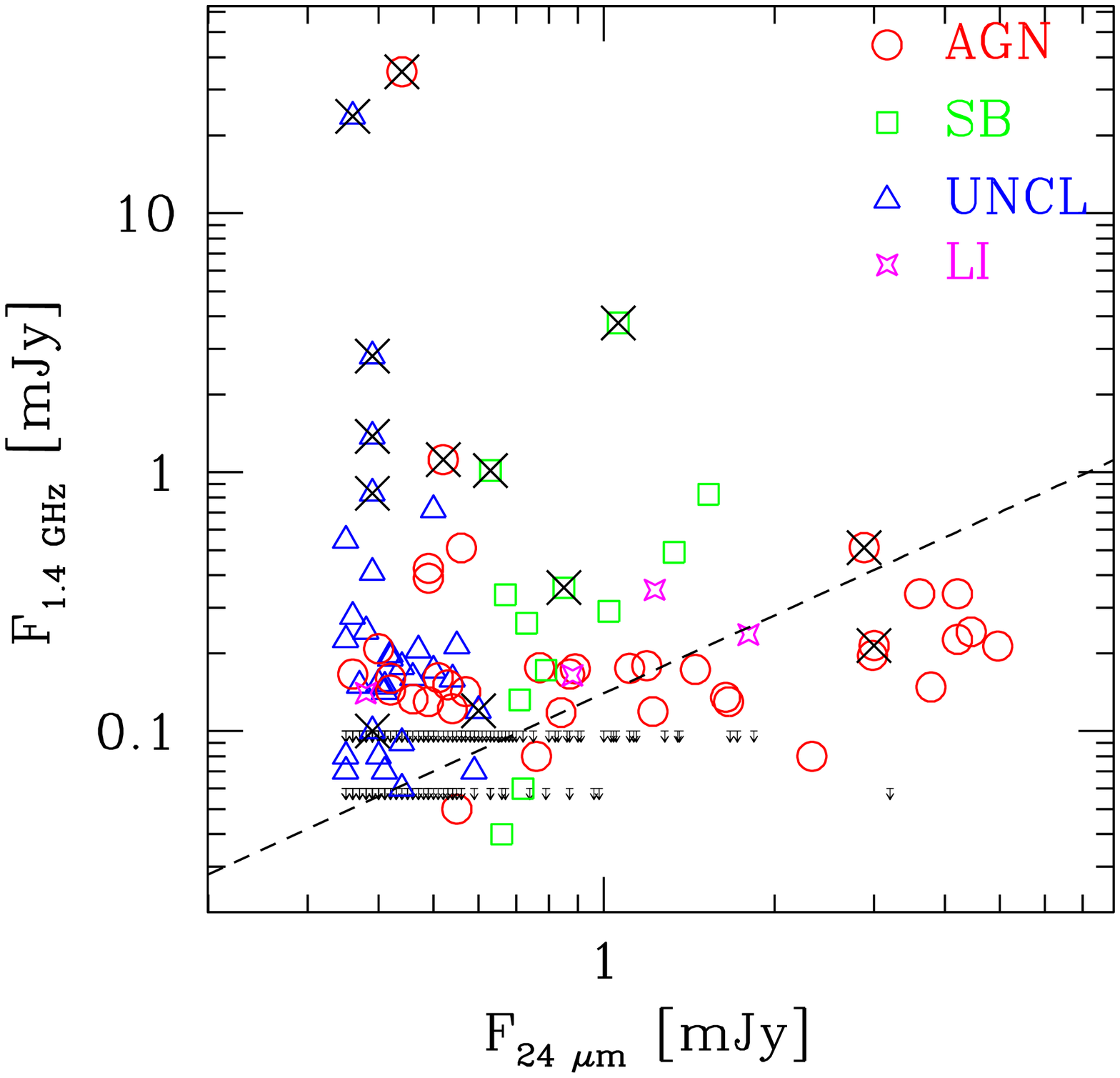}
\includegraphics[width=8.0cm]{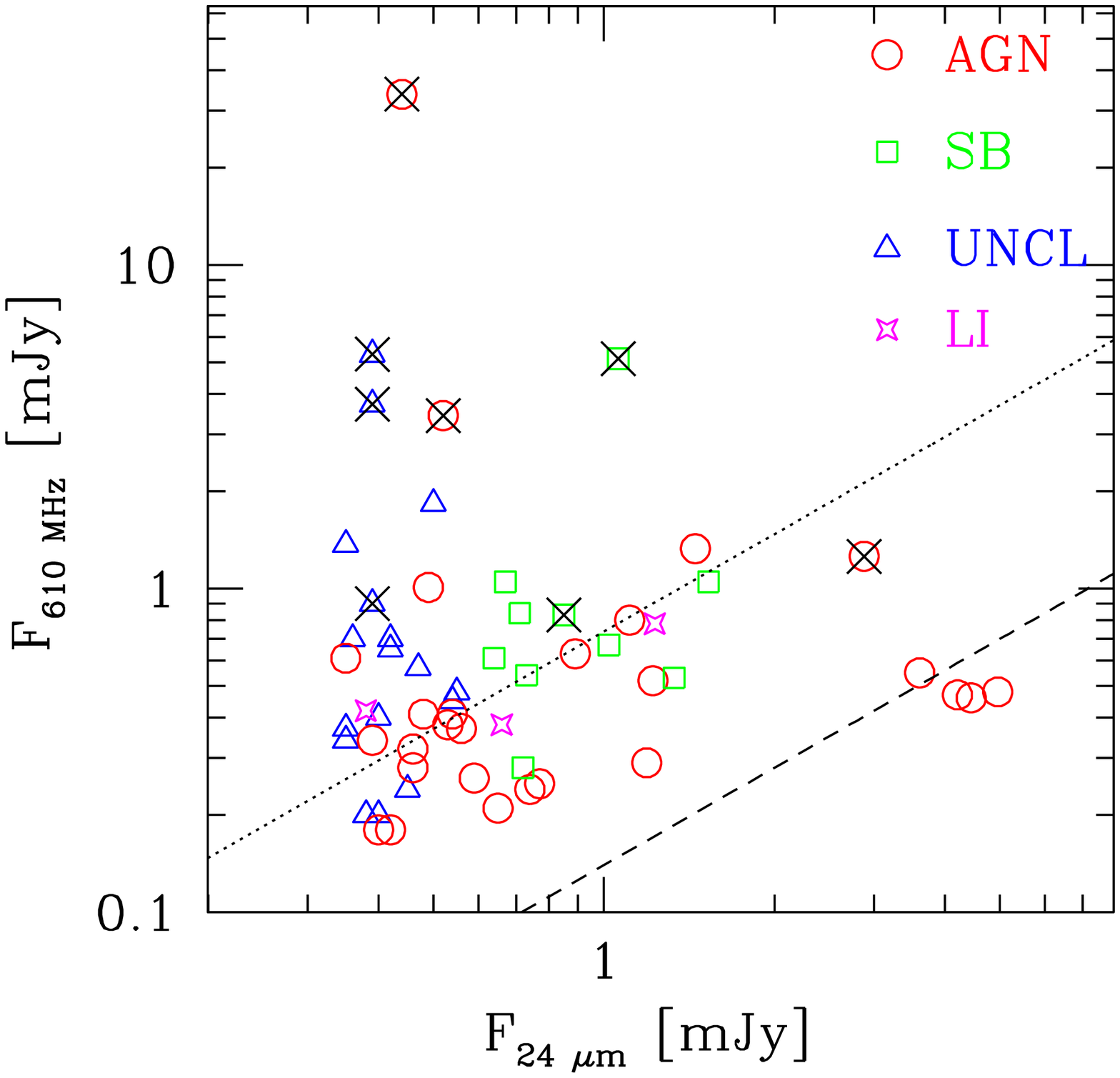}
\caption{Left-hand panel: 1.4~GHz vs 24$\mu$m fluxes for optically obscured 
{\it Spitzer} sources. (Red) circles are for objects classified as AGN, 
(green) squares for candidate star-forming galaxies, (blue) triangles for 
unclassified sources and (magenta) stars for low-z interlopers. Crosses indicate 
those objects whose radio activity is most likely associated to a radio-loud AGN 
(see text for details). The upper set of arrows represents the radio 
upper limits for those {\it Spitzer} sources in the Magliocchetti et al. 
(2007) sample without a radio counterpart in the Condon et al. (2003) 
catalogue, while the lower set of arrows corresponds to sources without a 1.4 GHz 
counterpart also in the Morganti et al. (2004) dataset.
The dashed line is the best fit to the Appleton et al. (2004) data.
Right-hand panel: as before but for 610~MHz radio fluxes. 
The dashed line still represents the Appleton et al. (2004) results, while 
the dotted line is the Appleton et al. (2004) best-fit as converted to 610 MHz 
fluxes by assuming an average radio spectral index for the sources 
$\langle \alpha\rangle =2$.
\label{fig:frvsf24}}
\end{figure*}

The slopes of the radio counts of optically obscured {\it Spitzer} 
sources at both 1.4~GHz and 610~MHz are in remarkable agreement with those 
found for the parent radio catalogues of Morganti et al. (2004) and Garn 
et al. (2007) (see \S2 and Figures~\ref{fig:morganti} and \ref{fig:GMRT}). 
If one again excludes some kind of cosmic conspiracy, this finding suggests 
that the class of radio-emitting objects set at 
redshifts $z\sim 2$ and with an enhanced mid-IR emission might not be too 
dissimilar from the population which dominates the (global) faint radio 
counts. 

A direct comparison between the amplitudes $A_S$ and $A_S^\prime$ as obtained 
above for the radio counts of optically obscured sources at 1.4~GHz and 
610~MHz and renormalized so to account for the different flux intervals 
adopted in estimating $\Delta N/\Delta F$ at the two radio frequencies 
($\Delta F_{1.4\rm GHz}=0.1$~mJy in one case and 
$\Delta F_{610\rm  MHz}=0.2$~mJy  in the second case) and for the different 
total number of sources with a radio identification in the 1.4~GHz and 
610~MHz catalogues, shows that the amplitude as inferred from the 
610~MHz counts is about a factor 
two higher than that derived from the counts at 1.4~GHz ($156^{+26}_{-52}$ in 
the first case to be compared with $71^{+14}_{-21}$). If in the radio interval 
probed by the present observations we consider a SED for these sources 
which goes as a power-law of 
index $\alpha$ (the so-called {\it radio spectral index}, whereby we use 
the notation $F_R\propto \nu_R^{-\alpha}$, with $F_R$ radio flux and $\nu_R$ 
generic radio frequency), we can then see that such a factor two of 
discrepancy can only be reconciled if the sources present an average 
radio spectral index $\left<\alpha\right>\sim 1$. 
Values of the order of $\left<\alpha\right>=0$ are strongly disfavoured by 
the data and this implies that the majority of our radio-identified 
{\it Spitzer} sources cannot be made of 'classical' flat-spectrum AGN.

\subsection{Mid-IR vs radio emission}
In \S5.1, investigations of the radio number counts of 
$z\sim 2$, mid-IR-selected sources have shown that these objects are probably 
not too different from the population which dominates the total (relatively 
faint) radio counts at both 1.4~GHz and 610~MHz.    
Furthermore, we have found that it is likely that radio emission 
in these {\it Spitzer} sources originates from similar processes, regardless 
of whether the mid-IR SEDs identify them as AGN-dominated or powered by an 
intense event of stellar formation. Finally, we could exclude 'classical' 
flat-spectrum AGN as the typical object which constitutes our radio-identified 
sample, as a comparison between radio counts at 1.4~GHz and 610~MHz requires 
average radio spectral indices $\left<\alpha\right>\sim 1$.

\begin{figure*}
\includegraphics[width=8.0cm]{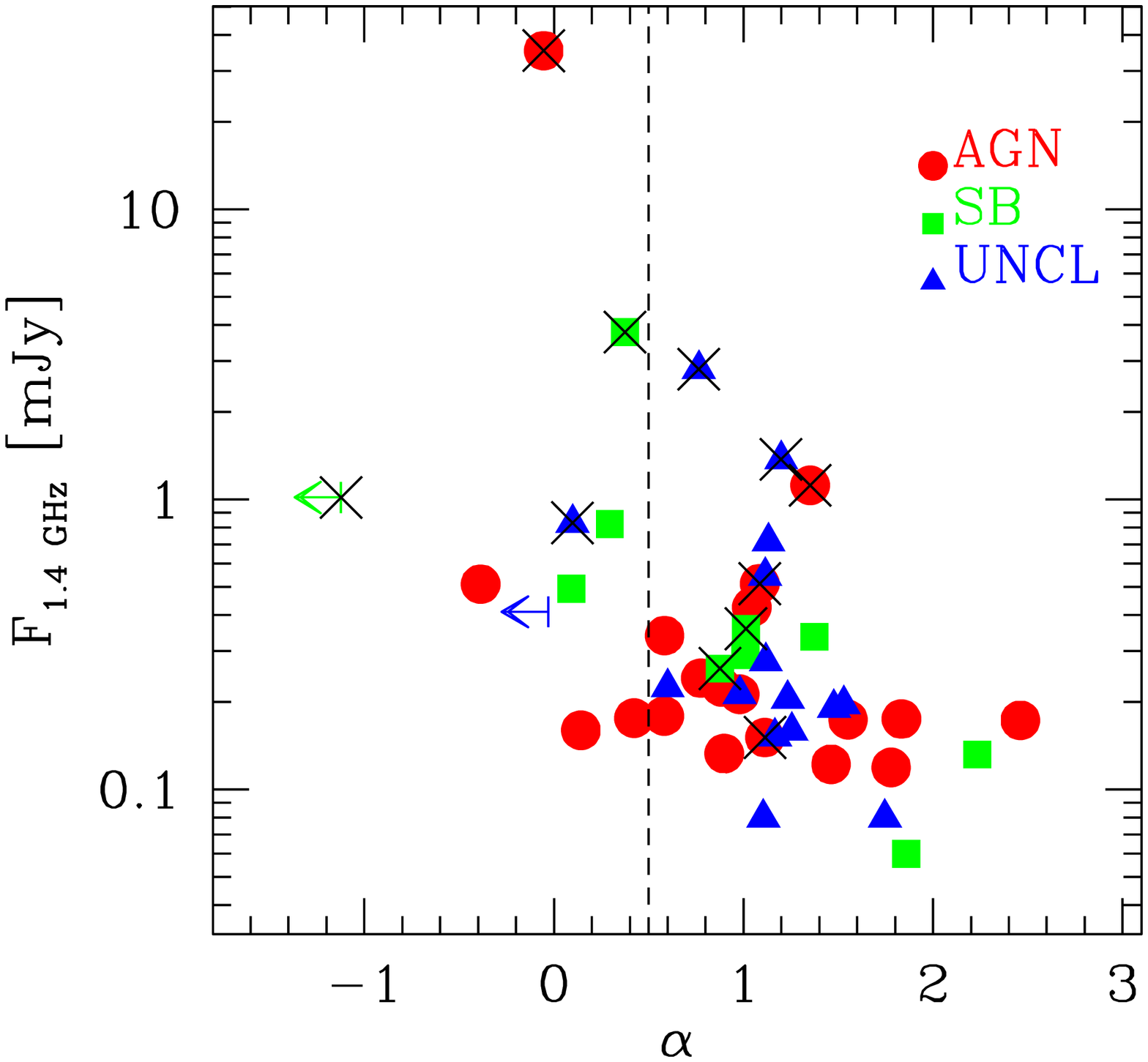}
\includegraphics[width=8.0cm]{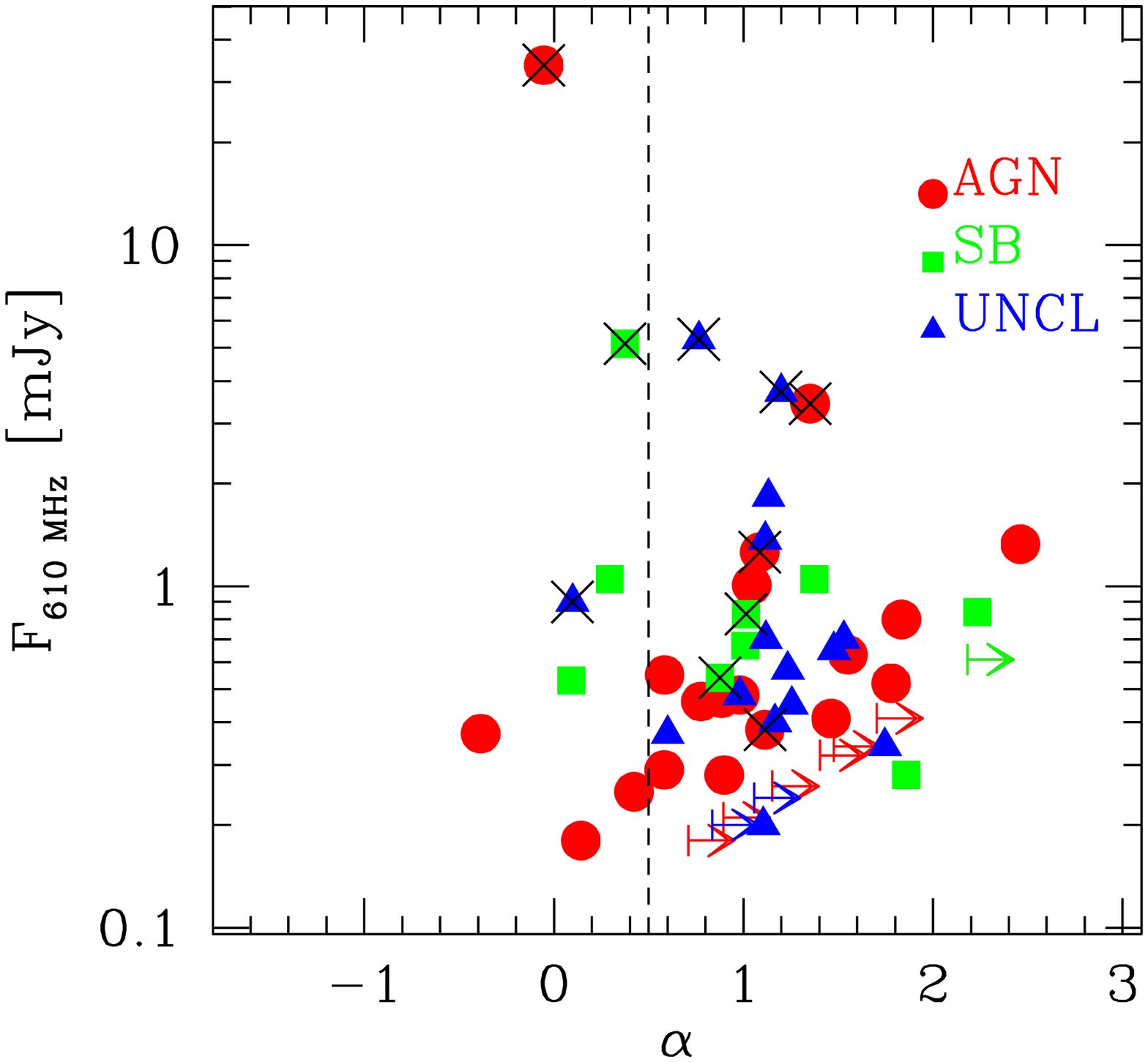}
\caption{Radio spectral index $\alpha$ as a function of 1.4~GHz 
(left-hand panel) and 610~MHz (right-hand panel) radio flux densities. 
Filled (red) dots are for the population of candidate AGN, (green) squares 
for star-forming galaxies, while (blue) triangles 
represent unclassified objects. Upper and lower limits are also indicated 
as colour-coded. Crosses indicate those sources whose radio activity 
is most likely associated to a radio-loud AGN 
(see text for details). The dashed lines in both panels mark the 
the transition between flat-spectrum 
($\alpha\simlt 0.5$) and steep-spectrum ($\alpha\simgt 0.5$) radio objects.
\label{fig:alpha_radio}}
\end{figure*}

More information on the nature of radio emission for the sources included in 
our sample can be obtained from a direct comparison of mid-IR and radio 
emission. A tight correlation between far-infrared and radio 
emission from galaxies has in fact been observed for over three decades, 
since the pioneering works of e.g. Condon et al. (1982) based on very small 
samples of galaxies. 
The origin of this correlation is thought to lie in the link between massive 
stars, which generate infrared emission by rehating dust, and supernovae, 
which accelerate cosmic rays that generate radio synchrotron radiation 
(see e.g. Harwit \& Pacini 1975; Condon 1992).

Observations with the InfraRed Astronomical 
Satellite (IRAS; Dickey \& Salpeter 1984; de Jong et al. 1985; Condon \& 
Broderik 1986) conclusively established such a correlation 
over a broad range of Hubble types and luminosities, from rich dwarfs to 
ultraluminous infrared galaxies at least in the local universe 
($z\simlt 0.1$). The advent of the Infrared Space Observatory (ISO) could for 
the first time probe the relatively higher-z universe, and it was found that 
even at 15$\mu$m there was a loose correlation between mid-IR emission 
and the radio continuum (e.g. Cohen et al. 2000; Gruppioni et al. 2003). 
More recently, works based 
on data obtained with the {\it Spitzer} Space Telescope have proved 
the IR-radio correlation in starforming galaxies to hold both at 70$\mu$m 
and -- although with a larger dispersion -- also at 24$\mu$m at least to 
redshifts $\sim 1$, and possibly to $z\sim 2$ if one relies on stacking methods  
(e.g. Appleton et al. 2004; Norris et al. 2006; Boyle et al. 2007). 

The 24$\mu$m flux as a function of radio flux both at 1.4~GHz and at 610~MHz 
for the sources in our sample is reported in Figure~\ref{fig:frvsf24}.
(Red) circles indicate those objects which are classified as AGN on the basis 
of their mid-IR photometry (where we put together the classes of AGN0 and 
AGN1, see \S4), while (green) squares represent candidate starburst 
galaxies and (blue) triangles unclassified sources. The (few) low-z interlopers 
are marked by (magenta) stars. We cross out those 
objects which are doubtful associations (i.e. J172217.4+601003) and also 
those sources (namely J171054.4+594426, J171143.9+600741, J172005.0+592430, 
J171527.1+585802, J172103.6+585052, J171427.8+592828, J171417.6+600531, 
J171312.0+600840, J172305.1+593841, J172256.4+590053, J171948.6+585133, 
J172258.9+593126, J171239.2+591350) 
which, on the basis of their morphology likely belong to the class of 
radio-loud AGN, not observed to follow the mid-IR/radio correlation 
(see the Appendix for more details on these objects). 
The dashed lines in both panels represent the relationship found by Appleton 
et al. (2004) ($q_{24}=-0.84\pm 0.28$) for their (uncorrected) data. 

With the exception of objects which are 
morphologically identified as radio-loud AGN, the overwhelming majority of 
optically obscured {\it Spitzer} sources with a radio counterpart either in the 
Condon et al. (2003) or in the Morganti et al. (2004) catalogues agree,  
although with a large scatter, with the Appleton et al. (2004) relation. 
Furthermore, a similar behaviour is also found for the upper radio limits 
of those sources which 
do not have a radio counterpart in either the Condon et al. (2003; 
upper set of arrows at $F_{1.4 \rm GHz}\simeq 0.1$~mJy in Figure~10) or Morganti et al. 
(2004; lower set of arrows at  $F_{1.4 \rm GHz}\simeq 0.06$~mJy in Figure 10) datasets.
This is rather surprising for two reasons. The first one is that the above relation 
is observed to hold for any type of galaxy, independent of their mid-IR 
classification, and therefore also for AGN. Although this result is not 
striking {\it per se} as also most Seyfert galaxies are shown to follow 
the IR-radio correlation (see e.g. Roy et al. 1998), this suggests that 
the 1.4~GHz luminosity of all the sources examined in this work is dominated 
by star formation activity, despite the presence of an AGN.
In this respect, the large scatter of our data around the Appleton et al. (2004) relation 
can be at least partially interpreted as due to a -- less important --
contribution of AGN origin, in general not observed to correlate with optical and/or 
IR emission (e.g. Andreani et al. 2003).\\
The second reason is instead intimately linked to the form of the SED of 
star-forming objects probed by {\it Spitzer} at redshift 
$z\sim [1.6-2.7]$. Such a redshift interval in fact samples the SED region 
which is dominated by strong silicate absorption and 
PAH emission lines (see Figure~\ref{fig:SED}), which make the observed
24$\mu$m luminosity of these sources extremely variable even if one moves 
in redshift by a very small amount. Such an extreme variability most likely constitutes 
the dominant cause for the observed spread around the Appleton et al. (2004) relation.
We note that a very similar result on the relation between 1.4~GHz and 24$\mu$m 
fluxes was obtained by Weedman et al. (2006), even though for a much smaller sample 
of $z\sim 2$, mid-IR bright objects (both candidate AGN and starbursts) in the FLS. 

When compared with what was found at 1.4~GHz, the sources in our sample 
seem to show a more loose trend between radio emission at 610~MHz and 
mid-IR fluxes. The right-hand panel of Figure~\ref{fig:frvsf24} visualizes 
this result and also indicates that, if anything, some kind of a correlation 
similar to that found by Appleton et al. (2004) could only be envisaged if 
our population of radio-identified {\it Spitzer} sources present an average 
radio spectral index $\left<\alpha\right>\simgt 1$ 
(the value $\left<\alpha\right>=2$ chosen for the 
dotted line in Figure~\ref{fig:frvsf24} is simply to guide the reader's eye), 
in agreement with what found in \S5.1 from an investigation of 
the radio number counts for these sources. We will investigate this issue 
in greater detail in the next Section.

\section{The radio spectral index}
\begin{figure}
\includegraphics[width=8.0cm]{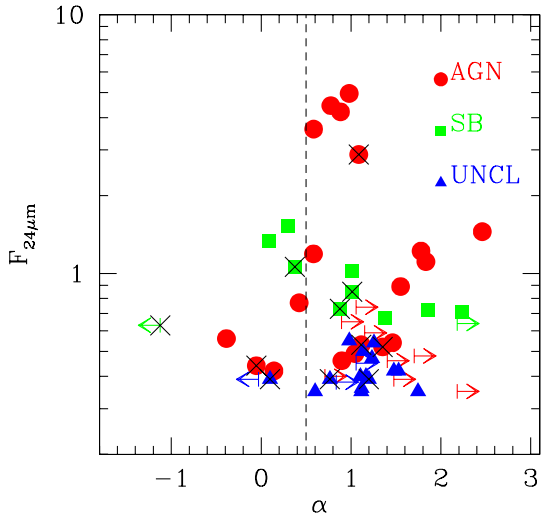}
\caption{Radio spectral index $\alpha$ as a function of $F_{24\mu \rm m}$ 
flux density. Filled (red) dots are for AGN, (green) squares for 
star-forming galaxies, while (blue) triangles 
are for unclassified objects. Upper and lower limits are also indicated as 
colour-coded. Crosses indicate those sources whose radio activity 
is most likely associated to a radio-loud AGN 
(see text for details). The dashed line marks 
the transition between flat-spectrum and steep-spectrum radio objects 
($\alpha\sim 0.5$).
\label{alpha_24}}
\end{figure}

\begin{figure}
\includegraphics[width=8.0cm]{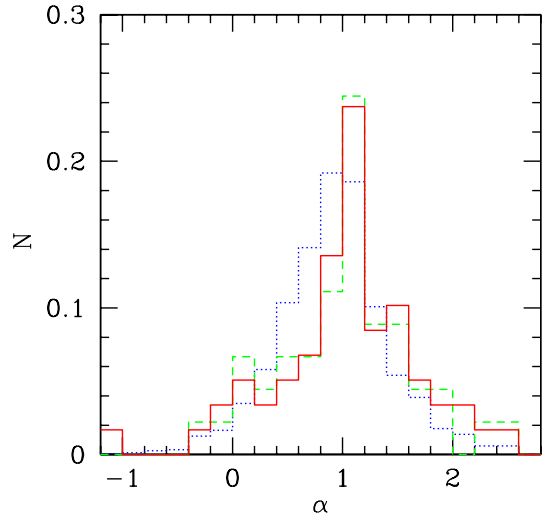}
\caption{Fractional distribution of radio spectral indices $\alpha$. The dotted (blue) 
line corresponds to all faint radio sources identified at both 610~MHz and 1.4~GHz (1516 objects), 
the dashed (red) line is for 
the subsample of $z\sim 2$ {\it Spitzer}-selected sources (43 objects), while the solid (green) 
line represents the result obtained by also including in the 
$z\sim 2$ {\it Spitzer} sample objects with estimated upper and lower limits on $\alpha$ (56 objects).
\label{fig:alpha_hist}}
\end{figure}

\begin{figure*}
\includegraphics[width=8.0cm]{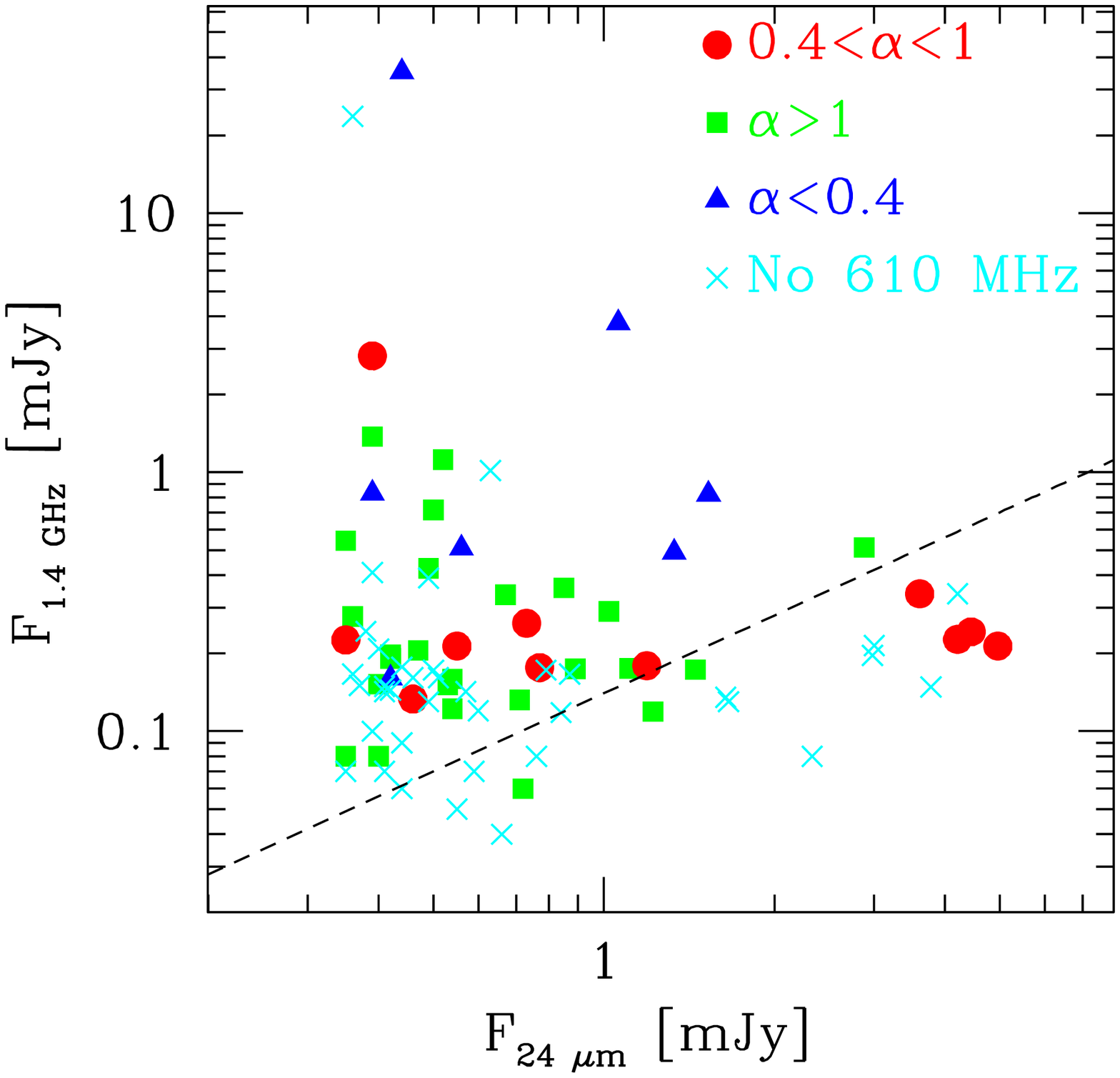}
\includegraphics[width=8.0cm]{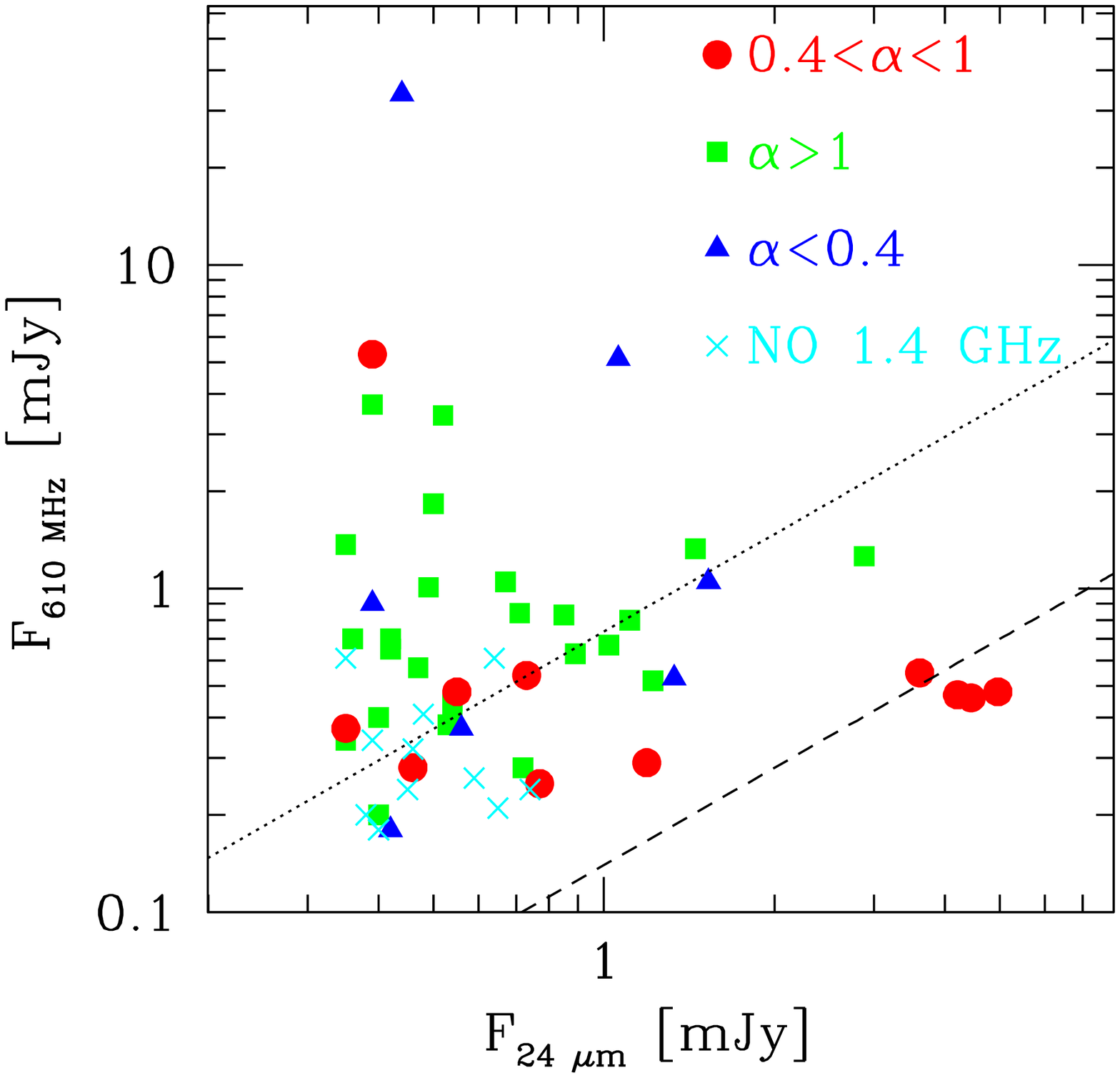}
\caption{24$\mu$m vs 1.4~GHz (left-hand side) and 610~MHz (right-hand side) 
fluxes for objects with different radio spectral index $\alpha$. Crosses in 
the left-hand panel identify those sources which do not have a 610~MHz 
counterpart, while those on the right-hand panel are for objects without 
1.4~GHz fluxes above the detection thresholds (see text for details). The line coding is as in 
Figure~10.}
\end{figure*}

While the analysis performed in \S5.1 and \S5.2 could only provide 
information on the average radio spectral properties of the population of 
radio-identified {\it Spitzer} sources set at $z\sim 2$, a direct 
estimate of the 610~MHz-to-1.4~GHz radio spectral index $\alpha$ is 
possible for all those sources which present detected radio fluxes at both 
the investigated frequencies. 

Under the assumption of a power-law behaviour for the SED in the radio 
interval between 610~MHz and 1.4~GHz, $\alpha$ was then calculated for 
45 objects belonging to our sample. 
12 lower limits and 3 upper limits have been furtherly added to our list 
of estimated $\alpha$'s as explained here. Since the Condon et al. (2003) survey is 
deeper than the FLS GMRT of Garn et al. (2007) ($\sim 100 \mu$Jy vs 
$\sim 200\mu$Jy; see \S2.2), we expect all the sources detected in the 
Garn et al. (2007) dataset to be also detected in the Condon et al. (2003) 
5$\sigma$ maps unless they have steep, $\alpha\simgt 0.5$ radio spectra. 
12 optically faint {\it Spitzer} sources have a radio counterpart in the GMRT 
only, and the lower limits for their radio spectral indices  
have been estimated as $\alpha_{\rm up}
={\rm log}_{10}(100[\mu \rm Jy]/F_{610\rm MHz}[\rm \mu Jy])/(-0.36)$, 
where -- as shown above -- 100$\mu$Jy approximately corresponds to 
the completeness limit probed by the Condon et al. (2003) survey.
Conversely, there are three sources which are relatively bright at 1.4~GHz, 
and for which one would have also expected a detection at 610~MHz unless 
endowed with inverted (i.e. negative values of $\alpha$) spectra. For these 
sources with $F_{1.4\rm GHz}\ge 400\mu$Jy (corresponding to the limit for 
completeness of the GMRT survey in the case of $\alpha$=0), the lower limits 
in their radio spectral indices were instead estimated as  $\alpha_{\rm low}
={\rm log}_{10}(F_{1.4\rm GHz}[\rm \mu Jy]/400[\mu\rm Jy])/(-0.36)$.  
All the remaining sources have 1.4~GHz fluxes which are faint enough to 
account for non-detection at 610~MHz. No estimate of $\alpha$ was possible 
for these objects. 
 
Values for the radio spectral index for the sources in our sample 
estimated as above are given in Table~1. Figures~\ref{fig:alpha_radio} and 
\ref{alpha_24} show their trend as a function respectively of 1.4~GHz, 
610~MHz and 24$\mu$m fluxes for the three sub-classes of AGN (red circles), 
star-forming galaxies (green squares) and unclassified objects (blue 
triangles). As it was in \S5.2, sources with radio morphologies which suggest 
the presence of a radio-loud AGN have been excluded from our analysis and have 
been marked with a cross in the various plots.

Two main features can be gathered from the investigation of 
Figures~\ref{fig:alpha_radio} and \ref{alpha_24}. The first one is that there 
is no relationship between the distribution of $\alpha$ values and either 
radio/mid-IR fluxes. The second, intriguing finding is that there are very 
few flat-spectrum sources with $\alpha< 0.5$, in agreement with the 
results of \S5.1 and \S5.2. The overwhelming majority of our objects (36 if 
one also includes the lower limits, value which becomes 34 by removing the low-z interlopers) 
instead presents very steep, $\alpha\ge 1$ 
spectral indices, some of them even reaching values of $\alpha\sim 2.5$.

Giving the limiting fluxes of the Condon et al. (2003) and Garn et al. (2007) surveys, 
the distribution of radio spectral indices as obtained by matching these two datasets will 
necessarily be biased towards high, $\alpha\simgt 0.5$, figures. Therefore, in order to assess 
the statistical significance of our findings, we have compared the distribution of $\alpha$ 
values for {\it all} the sources which appear in both the GMRT and Condon et al. (2003) 
catalogues (1516 objects to a separation between 610~MHz and 1.4~GHz positions 
$d=6^{\prime\prime}$) with the distribution obtained for our sample of $z\sim 2$ radio-detected 
{\it Spitzer} galaxies. 
As Figure~13 shows, although in general not too different from that of the totality of radio sources 
(a KS test reports 
values of 0.99 and 0.98 respectively if one considers radio detections only or also takes into account 
upper and lower limits on $\alpha$), the distribution of $\alpha$ in the case of {\it Spitzer} galaxies
(dashed -- red --
line) is systematically shifted towards higher values of the radio spectral index. This 
trend becomes even stronger if one also includes in the analysis the upper and lower limits on 
$\alpha$ (solid -- green -- histogram in Figure~13), estimated as discussed above. 
The statistical significance of this high-$\alpha$ tail is quite high:  for the whole population of 
faint radio sources with identifications at both 610~MHz and 1.4~GHz we have that the percentage 
of sources with $\alpha\simgt 1$ is 43\%, while this number rises to 60\% (i.e. 26 galaxies out 
of 43 or 34 out of 56 by also including upper and lower limits on $\alpha$, see Table~1) for the 
sample of $z\sim 2$ {\it Spitzer} galaxies.

This result is quite surprising. Very steep radio spectral indices  
tend to identify the population of Ultra Steep Spectrum sources (USS), mostly 
radio-loud galaxies set at substantial redshifts. Indeed, objects 
with high values of $\alpha$ are typically investigated to look for very 
high redshift radio 
galaxies (e.g. De Breuck et al. 2000; 2002). However, the objects in our 
sample are not 'a priori' radio-loud AGN as they have been merely selected 
as high-z, strong mid-IR emitters. Furthermore, we find that the majority of our 
sources present values of $\alpha> 1$ independent of whether their 
mid-IR emission
is indicative of a starburst-dominated galaxy or of obscured AGN  activity.

Taking at face value our results, 
one would then conclude that the overwhelming majority of the sources in our 
radio-identified sample is made of radio-loud AGN. This is however in 
striking disagreement with the findings of \S5.2 which show that most 
of the very same objects follow a 24$\mu$m-to-1.4~GHz relation between fluxes 
which closely resembles that typical of starburst-powered galaxies.
Indeed, if one reproduces the $F_{24\mu\rm m}$--$F_{1.4\rm GHz}$ 
plot, this time by grouping together sources with similar values of 
$\alpha$ (where the three broad classes are $\alpha<0.4$, flat-spectrum AGN; 
$0.4\le \alpha\le 1$, possible starbursts; $\alpha>1$ USS; see Figure~14), it is clear that, 
while all but one of the objects with $0.4\le\alpha\le 1$  do fall in the 
allowed region identified by the Appleton et al. (2004) 
relation, the same can be said for the majority of sources with 
$\alpha> 1$. It is however comforting that most flat-spectrum sources in our sample -- 
in many cases coinciding with those objects which present a radio morphology 
typical of a radio-loud AGN  -- do instead lie outside the 'allowed' range for 
radio emission from star formation activity. 

A natural explanation which can reconcile the above discrepant findings 
is provided by envisaging a 
two-component radio spectrum, where the flatter $\alpha_{\rm SF}\sim 0.8$ 
component originates from processes associated to star-formation, while the 
steeper $\alpha_{\rm AGN}\sim2$ one is due to the presence of an AGN. 
In such a two-component model, the 1.4~GHz emission would primarily stem 
from star formation, while the 610~MHz signal could mainly be attributed to 
the radio-loud AGN.  This framework could then also explain why the 1.4~GHz 
emission in these sources follows that of star forming systems, while the relation is more 
loose in the case of 610~MHz and $24\mu\rm m$ fluxes (see 
Figure~\ref{fig:frvsf24}). 
We note that, an implicit but very important implication of the above 
discussion is of a cohabitation of the two processes of star formation and 
accretion onto an AGN, both expected to take place at the same time within the same systems.

A number of works can be found in the literature which try to explain 
 the presence of systems with very steep spectral slopes at high redshifts. 
For instance, by comparing the extremely steep 
spectral index sources associated with galaxies residing closest to the 
cluster centres, Klamer et al. (2006) found 
that steeper spectra can be explained by pressure-confined radio 
lobes which have slow adiabatic expansion losses in high-density 
environments. Alternatively, one can attribute the steepening of the radio 
spectrum at low frequencies as due to the scattering between CMB photons and  
relativistic electrons at $z\sim 2$ where the CMB energy density is 
significantly higher than it is at later epochs 
(e.g. Martinez-Sansigre et al. 2006). 
Both theories need high-z radio sources to reside 
in very dense environments. This is in agreement with the results of 
Magliocchetti et al. (2007; 2007a) which find the parent population 
of these radio-identified {\it Spitzer} sources 
to be hosted by very massive/cluster-like structures. 
Furthermore, recent clustering studies performed on the class of USS 
also find the latter objects to reside in $M\simgt 10^{13.4} M_\odot$ systems 
(Bornancini et al. 2006). 
The remarkable similarity between the Magliocchetti et al. (2007; 2007a) and 
Bornancini et al. (2006) results indicate that USS and $z\sim 2$ {\it Spitzer}-selected sources 
reside in very similar environments, finding which strengthens the case for a relationship between 
these two populations.


\section{Conclusions}
This paper has presented an analysis of the radio properties
of a subsample of optically faint ($R\simgt 25.5$), 24$\mu$m-selected 
galaxies observed by {\it Spitzer} in the FLS (Magliocchetti et al. 
2007). These objects have been cross-correlated with a number of radio 
catalogues which cover the same region of the sky, namely that of Condon et 
al. (2003) which probes 1.4~GHz fluxes brighter than $\sim 100 \mu$Jy, that 
of Garn et al. (2007) -- which probes 610~MHz fluxes brighter than 
$\sim 200 \mu$Jy -- and, on a smaller portion of the sky, that of Morganti et 
al. (2004) which reaches 1.4~GHz fluxes as faint as $\sim 40\mu$Jy. 

70 optically faint {\it Spitzer} sources have been identified in the Condon 
et al. (2003) catalogue, 33 in the Morganti et al. (2004) dataset, while 52 
are found in the survey performed by Garn et al. (2007). After performing
a number of corrections to account for multiple identifications, 
sources erroneously split in the original 
{\it Spitzer} catalogue into different components and mid-IR objects with real radio 
counterparts at one of the two radio frequencies which were further away than the allowed 
($10^{\prime\prime}$) matching radius, we end up with a sample 
of 96 radio-identified, optically faint, mid-IR emitting sources, 45 of which have an 
identification at both 1.4~GHz and 610~MHz. The fraction of radio identifications 
is a strong function of 24$\mu$m flux: almost 
all sources brighter than $F_{24\mu\rm m}\sim 2$~mJy are endowed with a 
radio flux at both 1.4~GHz and 610~MHz, while this fraction drastically 
decreases by lowering the flux level.

IRAC photometry for all those sources which also have detected fluxes in at 
least one of the four 8$\mu$m, 5.8$\mu$m, 4.5$\mu$m and 3.6$\mu$m channels 
(64 out of 96), allows to classify them
into two categories: obscured AGN (45 sources) and systems mainly 
powered by starformation activity, (SB, 13 objects). We also find five low-z 
(i.e. $z\simlt 0.5$ m51-type) interlopers, while the remaining 33 sources are 
unclassified. Furthermore, with the help of IRAC photometry it was 
possible to assign broad redshift intervals to all those sources (mostly AGN) which 
presented in the lowest 3.6$\mu$m and $4.5\mu$m wavelength channels of IRAC
a 'bump' compatible with being produced by an evolved (old) stellar 
population. The majority ($\sim 66$\%) of these galaxies 
reside at redshifts $z\simgt 1$, in agreement with other studies mainly 
based on mid-IR and near-IR spectroscopy of optically faint, 24$\mu$m-selected galaxies 
(i.e. Weedman et al. 2006; Yan et al. 2005; 2007; Brand et al. 2007). We stress that this  
inferred fraction can only be considered as a lower limit to the real portion of faint 
{\it Spitzer} sources set at high redshifts. In fact, because of their characteristic spectral 
properties in the mid-IR regime, we expect the majority of star forming systems (too 
faint at the IRAC frequencies to have measurable 3.6$\mu$m and/or 4.5$\mu$m fluxes) to be 
indeed located in the $z\sim [1.6-2.7]$ redshift range.
 
A small fraction of objects in our sample present radio morphologies such as 
jets and/or lobes compatible with them being 
identified as radio-loud AGNs. Interestingly enough, we find that for most of these few extended 
radio sources the mid-IR emission is associated to such peripheral regions rather than steming 
from the centre of radio activity, generally coinciding with the location of the AGN. However,
the majority of the objects in our sample present unresolved radio images.

A compared analysis of the radio number counts for optically obscured 
{\it Spitzer} sources indicates that the $\Delta N/\Delta F$ as estimated at 1.4~GHz can 
only be reconciled with what found at 610~MHz if the 
population under investigation is endowed with an average value for the radio spectral index  
(defined as $F_R\propto \nu_R^{-\alpha}$, where $F_R$ is the radio flux 
and $\nu_R$ the generic radio frequency) $\left<\alpha\right>\simgt 1$. 
Classical, 'flat spectrum' radio sources can confidently be excluded 
as the typical objects constituting our sample. Furthermore, 
we have found that the radio number counts of sources classified as AGN and of those 
identified as starburst galaxies are quite similar, evidence which suggests 
that radio emission in $z\sim 2$ {\it Spitzer} galaxies originates from similar process(es), 
despite of the different mid-IR emission.

Direct investigations of the relation between 24$\mu$m and 1.4~GHz fluxes 
show that the overwhelming majority of those galaxies not excluded from our 
analysis because morphologically classified as radio-loud AGN follow the 
relationship 
identified for $0\simlt z\simlt 1$ star-forming objects by Appleton et al. 
(2004), although with a large scatter. This happens regardless of whether 
the galaxy has been classified as an AGN or a starforming system on the basis 
of its mid-IR colours. The distribution of 24$\mu$m vs 610~MHz fluxes is 
instead found to be more scattered.

The majority of these radio-identified objects 
(26, a figure which rises to 34 if one also includes 
sources with estimated lower limits on $\alpha$) present very steep, $\alpha> 1$ 
radio spectral indices, some galaxies being endowed with $\alpha$'s 
as high as 2.5. This excess of galaxies with large $\alpha$ values is statistically 
significant as it corresponds to 60 per cent of our sample, to be compared 
with the 43 per cent found by considering {\it the whole population} of faint 
radio objects with an identification at both 610~MHz and 1.4~GHz.\\
Such very high figures for $\alpha$ would identify the corresponding 
sources as 
Ultra Steep Spectrum galaxies, generally high redshift radio-loud AGN. 
However this is in striking disagreement with what found for the relation 
between 24$\mu$m and 1.4~GHz fluxes for the very same objects, relation which 
would explain their (1.4~GHz) radio emission as mainly due to processes connected 
with star forming activity.

A natural explanation to the above issues could be found by assuming that 
AGN and star-formation activity are concomitant in the majority of 
$z\sim 2$, {\it Spitzer} sources, at least in those which present enhanced 
radio emission. 
The radio signal steming from these systems would then simply be the 
combination of two components: a shallower one -- dominating the spectrum  
at 1.4~GHz -- due to processes connected with star formation, and a steeper 
one -- being responsible for most of the 610~MHz signal -- connected with 
AGN activity. This framework could then also explain why the 1.4~GHz 
emission in our sources follows that of star forming systems, while the same 
does not seem to happen in the case of 610~MHz fluxes. 

Various explanations can be found in the literature to account for the presence of USS: 
from slow adiabatic expansion losses in high-density environments (e.g. Klamer et al. 2006) 
to the scattering between CMB photons and relativistic electrons at $z\sim 2$ 
(e.g. Martinez-Sansigre et al. 2006). However, the results presented in this work might provide
an alternative scenario. In fact, they suggest that high values for $\alpha$ 
might be due to the concomitant presence within the same systems of an AGN and of 
a star forming region: the AGN expansion would then be halted by the encounter with 
the cooler/denser sites in which star formation takes place. 
This would determine the 'strangling' of the AGN, causing its radio spectrum at low radio 
frequencies to steepen. Clearly, more theoretical work is needed in order to quantify the 
above issue and we are planning to present it in a forthcoming paper. For the time being, 
we note that intense star forming activity within a high redshift galaxy 
host of a USS has been recently reported by Hatch et al. (2007). 

From a more observational point of view, the 'ultimate truth' on these sources could only 
come from very high resolution (and, given their faintness, sensitivity) measurements, capable 
to clearly disentangle the emission related to the AGN 
to that associated to star forming regions. The advent of instruments such as ALMA 
will then provide the answers we need.

\section*{ACKNOWLEDGMENTS}
We thank G. De Zotti, R. Laing, C. De Breuck and our referee P. Appleton 
for very interesting discussions and comments that greatly helped shaping up the paper.

\section{Appendix}

\begin{figure*}
\includegraphics[width=18.0cm]{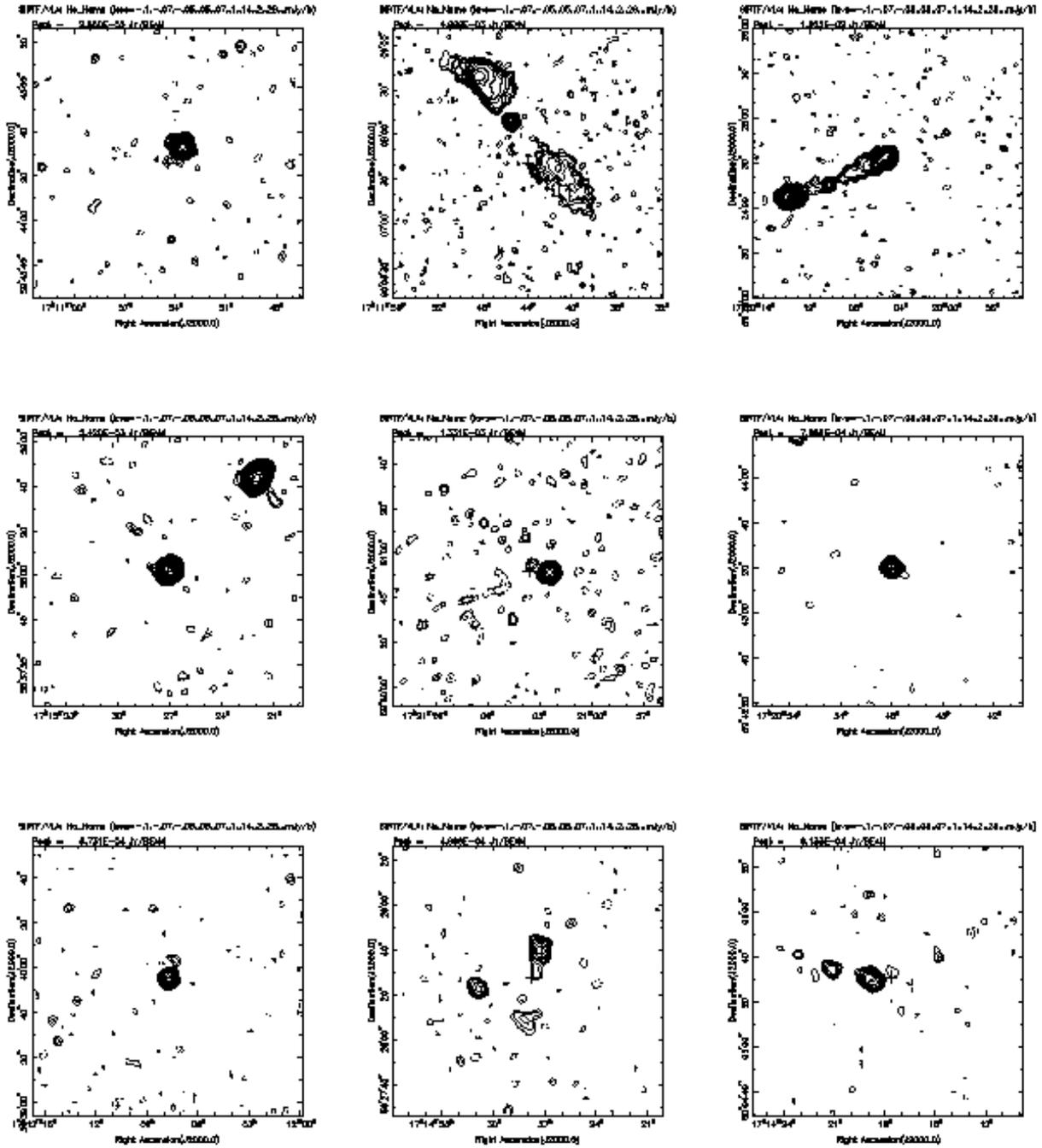}
\caption{Radio maps for all but one of the sources
of our sample brighter than $F_{1.4 \rm GHz}=0.8$~mJy. The centre of each
postage stamp (highlighted by a cross) corresponds to the position of
24$\mu$m emission. More detailed information on these objects
is provided in the Appendix. Except for the two postage stamps at the
top-centre and top-left panels which already appear in the Condon et al.
(2003) paper, the radio data are un-published and come from the Condon et al.
(2003) online FLS catalogue (available at $http://www.cv.nrao.edu/sirtf/$)}.
\label{fig:bright}
\end{figure*}


\begin{figure*}
\includegraphics[width=18.0cm]{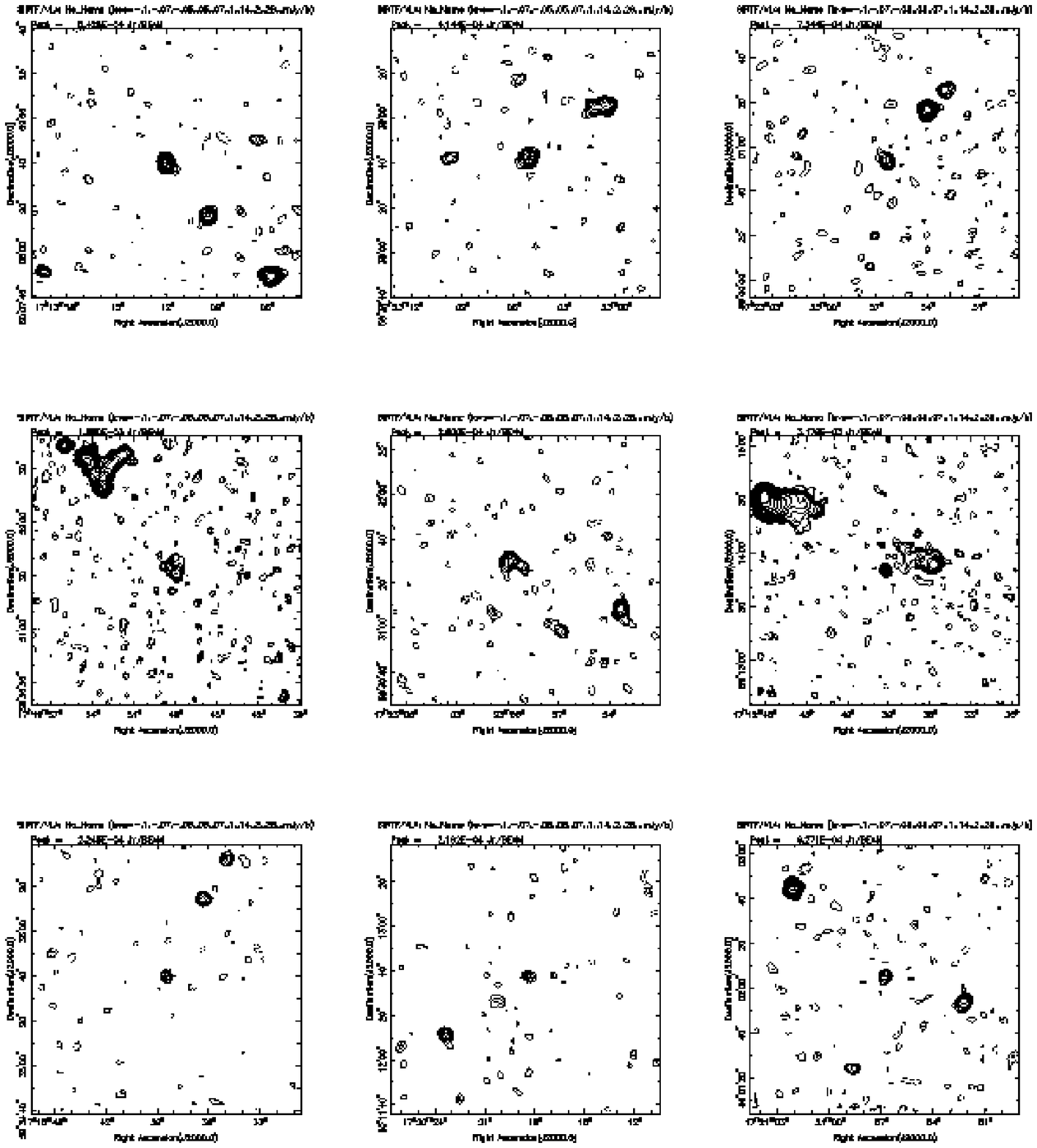}
\caption{Radio maps for all those radio sources
fainter than $F_{1.4 \rm GHz}=0.8$~mJy which present possible
extended/multi-component radio emission. The centre of each postage stamp
(highlighted by a cross) corresponds to the position of 24$\mu$m emission.
More detailed information on these objects is provided in the Appendix.
Radio data from the Condon et al. (2003) online FLS catalogue
(available at $http://www.cv.nrao.edu/sirtf/$).
\label{fig:extended}}
\end{figure*}


This Section provides some more detailed information and images for the most 
interesting sources found in our work. Comments on the likeliness of 
multi-component associations 
are mainly based on the method introduced by Magliocchetti et al. (1998).  
\begin{itemize}
\item J171239.2+591350. Possibly a triple radio source whereby the 24$\mu$m 
signal stems from 
the centre of radio emission (see middle panel in the last column 
of Figure~\ref{fig:extended}).
\item J172258.9+59312 and J172259.9+593129. Extended radio source made by two 
distinct components (see middle panel in the second column of 
Figure~\ref{fig:extended}) both at 24$\mu$m and at 1.4~GHz.  
The 610~MHz emission is instead only observed at the centre of the major 
component. Possibly a member of a double system.
\item J172256.4+590053. Possible triple radio source whereby the 24$\mu$m 
emission originates from one of the lobes (see upper panel in the last 
column of Figure~\ref{fig:extended}).
\item J171312.0+600840. Possible triple radio source whereby the 24$\mu$m 
emission originates from one of the lobes (see upper panel in the first 
column of Figure~\ref{fig:extended}).
\item J172018.1+592902. Single 24$\mu$m source split in the {\it Spitzer} 
catalogue into two components 
(J172018.1+592902 and J172018.1+592859). The 24$\mu$m flux reported in 
Table~1 is the sum of those of the two sub-components.
\item  J171527.1+585802. Possible double radio source, with the 24$\mu$m 
emission originating from one of the lobes (see middle panel in the first 
column of Figure~\ref{fig:bright}). Bright object both 
in the radio and mid-IR bands. 
\item J172305.1+593841. Possible double radio source, with the 24$\mu$m 
emission originating from one of the lobes (see upper panel in the second 
column of Figure~\ref{fig:extended}).
\item J171948.6+585133. Extended radio source (see middle panel in the first 
column of Figure~\ref{fig:extended}).
\item J171538.3+593540. Possible (but not likely) triple radio source with 
24$\mu$m radio emission originating from 
one of the lobes (see bottom panel in the first column of 
Figure~\ref{fig:extended}).
\item J172056.6+590206. Possible (but not likely) triple radio source with 
24$\mu$m emission coinciding with the 
centre for radio emission (see bottom panel in the last column of 
Figure~\ref{fig:extended}).
\item J172018.5+601237. Possible (but not likely) triple radio source with 
24$\mu$m radio emission originating from 
one of the lobes (see bottom panel in the second column of 
Figure~\ref{fig:extended}).
\item J171628.4+601342. Single 24$\mu$m source split in the {\it Spitzer} 
catalogue into 
three components (J171628.4+601342, J171628.4+601334, J171627.2+601342). 
The 24$\mu$m flux reported in Table~1 is the sum of those of the three sub-components.
\item J172353.2+601354. Single 24$\mu$m source split in the {\it Spitzer} catalogue into 
two components (J172353.2+601354 and J172353.2+601351). 
The 24$\mu$m flux reported in Table~1 is the sum of those of the two sub-components.
\item J171054.4+594426. Bright radio source. The 24$\mu$m emission originates 
from a lateral radio blob, possibly a jet (see upper panel in the first 
column of Figure~\ref{fig:bright}).
\item J171143.9+600741. Very bright triple source, which appears in the 
Condon (2003) catalogue as split into 7 components. The 24$\mu$m emission 
originates from one of the lobes (see upper panel in the second column of 
Figure~\ref{fig:bright}). Surprisingly enough, it does not have a 610~MHz 
counterpart within 30 arcsec from the centre of 24$\mu$m emission. 
\item J172005.0+592430. Very bright and extended triple component, which 
appear in the Condon (2003) catalogue as split into 5 components. The 
24$\mu$m emission originates from the centre of one of the lobes 
(see the upper panel in the last column of Figure~\ref{fig:bright}).
\item  J172103.6+585052. Close double system, whereby the 24$\mu$m originates 
from the smaller/fainter of the two radio components (see middle panel in the 
second column of Figure~\ref{fig:bright}).  
\item J171427.8+592828. Radio flux from the Morganti et al. (2004) catalogue. 
24$\mu$m emission associated to a lateral/jet-like blob. Possibly member of a 
double system (see bottom panel in the second column of Figure~\ref{fig:bright}).  
\item J171207.6+593956. Composite radio system formed by a bright central 
source which coincides with the centre of 24$\mu$m emission and a 
much fainter/smaller lateral blob, possibly a jet (see bottom panel 
in the first column of Figure~\ref{fig:bright}).                        
\item J172048.0+594320. Composite radio system formed by a bright central 
source which coincides with the centre of 24$\mu$m emission and a 
much fainter/smaller lateral blob, possibly a jet (see middle panel 
in the last column of Figure~\ref{fig:bright}). 
\item J171417.6+600531. 24$\mu$m emission from lateral and faint blob 
(see bottom panel in the last column of Figure~\ref{fig:bright}). 
No 610~MHz emission despite the relatively high 1.4~GHz flux.
%
%
\item  J172217.4+601003. 24$\mu$m emission far from radio image of closest 
1.4~GHz source. Likely a mis-identification.
\end{itemize}

\end{document}